\documentclass[a4paper,11pt]{article}

\usepackage{jheppub}
\usepackage{graphicx}
\usepackage{multirow}
\usepackage{amsmath,amssymb,amsfonts}
\usepackage{amsthm}
\usepackage{mathrsfs}
\usepackage[title]{appendix}
\usepackage{xcolor}
\usepackage{textcomp}
\usepackage{manyfoot}
\usepackage{booktabs}
\usepackage{algorithm}
\usepackage{algorithmicx}
\usepackage{algpseudocode}
\usepackage{listings}
\usepackage{enumitem}
\usepackage{tikz}
\usepackage[compat=1.1.0]{tikz-feynman}
\usetikzlibrary{shapes, arrows, calc}
\usetikzlibrary{decorations.pathreplacing, calligraphy}
\usepackage{colortbl}
\usepackage{bm}
\usepackage{cprotect}

\begin{document}

\begin{flushright}
    \preprint{KEK-TH-2633, DESY-24-094, YITP-24-76}
\end{flushright}

\title{Higgs boson production at $\mu^+ \mu^+$ colliders}

\author[a,b]{Yu Hamada,} 
\author[c,d]{Ryuichiro Kitano,} 
\author[c]{Ryutaro Matsudo,} 
\author[c]{Shohei Okawa,} 
\author[c,d]{Ryoto Takai,} 
\author[e]{Hiromasa Takaura} 
\author[c,d]{and Lukas Treuer} 

\affiliation[a]{Deutsches Elektronen-Synchrotron DESY, Notkestr. 85, 22607 Hamburg, Germany}
\affiliation[b]{Research and Education Center for Natural Sciences, Keio University, 4-1-1 Hiyoshi, Yokohama, Kanagawa 223-8521, Japan}
\affiliation[c]{KEK Theory Center, Tsukuba 305-0801, Japan}
\affiliation[d]{The Graduate University for Advanced Studies (SOKENDAI), Tsukuba 305-0801, Japan}
\affiliation[e]{Center for Gravitational Physics and Quantum Information, Yukawa Institute for Theoretical Physics, Kyoto University, Kyoto 606-8502, Japan\\}

\emailAdd{yu.hamada@desy.de}
\emailAdd{ryuichiro.kitano@kek.jp}
\emailAdd{matsudo@post.kek.jp}
\emailAdd{shohei.okawa@kek.jp}
\emailAdd{rtakai@post.kek.jp}
\emailAdd{hiromasa.takaura@yukawa.kyoto-u.ac.jp}
\emailAdd{ltreuer@post.kek.jp}

\abstract{We study Higgs boson production at $\mu^+ \mu^+$ colliders at high
energy. Since both initial-state particles are positively charged,
there is no $W$ boson fusion at the leading order, as it requires a $W^+ W^-$ pair. However, we find
that the cross section of the higher-order, $\gamma$- and $Z$-mediated
$W$ boson fusion process is large
at high center-of-mass energies $\sqrt s$,
growing as $(\log s)^3$. This is in contrast to the $\log s$ behavior of the leading-order $W$ boson fusion. Thus, even though it is a higher-order
process, the rate of Higgs boson production for 10~TeV energies at $\mu^+ \mu^+$ colliders with polarized beams can be as high
as about half of the one at $\mu^+ \mu^-$ colliders, assuming the same
integrated luminosity. To calculate the cross section of this process
accurately, we carefully treat the collinear emission of the photon in
the intermediate state. The thereby obtained large cross section
furthermore shows the significance of Higgs production with an extra
$W$ boson in the final state also at $\mu^+ \mu^-$ and $e^+ e^-$ colliders.}

\maketitle


\section{Introduction}

Higgs boson factories are the best-motivated future colliders,
enabling us to explore the nature of electroweak symmetry breaking in depth.
Among others, muon colliders at 10~TeV energies have been discussed as one
of the most attractive possibilities, since the size of the accelerator
facilities can be as compact as $\mathcal{O}(10)$~km in circumference,
and they can potentially probe physics up to scales as high as
$\mathcal{O}(100)$~TeV~\cite{Buttazzo:2020uzc, AlAli:2021let,
Black:2022cth}. By comparison, proton colliders with similar reach
require a circumference of at least $\mathcal{O}(100)$~km.

Recently, muon colliders based on ultra-slow muon technology~\cite{Nagamine:1995zz} have been
proposed~\cite{Hamada:2022mua}. This is because ultra-slow muons from laser-ionized
muonium, a  $\mu^+ e^-$ bound state, can be used to create $\mu^+$
beams with excellent emittances, i.e., very collimated beams that facilitate high luminosities. As such, the corresponding proposal
is to build on this technology, which was developed for the muon $g-2$/EDM
experiment at J-PARC~\cite{Abe:2019thb}.
Thus, under various assumptions for the parameters of the proposed collider facility, the luminosity is estimated to be $\sim 5.7 \times 10^{32}$~cm$^{-2}$s$^{-1}$ 
for a $\mu^+ \mu^+$ collider with center-of-mass energy 2~TeV, 
and $\sim 4.6 \times 10^{33}$~cm$^{-2}$s$^{-1}$ for a $\mu^+ e^-$ collider with center-of-mass energy 346~GeV~\cite{Hamada:2022mua}. 
For higher energies,
the improvement in the emittance as well as the 
larger boost factor further enhance the luminosity. In any case,
this estimate suggests that these colliders are realistic options for Higgs boson factories.

At such $\mu^+$-based colliders, even though there are no $s$-channel annihilation processes that are present in $\mu^+ \mu^-$ or $e^+ e^-$ colliders, Higgs boson production can still occur via vector boson fusion processes, which dominate over $s$-channel annihilations at energies beyond $\mathcal{O}(1)$~TeV. This is due to a logarithmic enhancement of the cross section for vector boson fusion as a function of center-of-mass energy.
At $\mu^+ e^-$ colliders, while $W$ boson fusion is possible, the center-of-mass energy is limited by the beam energy of the electron.
On the other hand, for $\mu^+ \mu^+$ colliders, one can consider, e.g., $\mathcal{O}(10)$~TeV
beam energies, but there is no $W$ boson fusion process at leading order since both of the
antimuons can only emit positively charged $W$ bosons. Although $Z$ boson fusion is possible
at leading order since it is independent of the muon charge, its cross section is about an
order of magnitude smaller than $W$ boson fusion at $\mu^+ \mu^-$ colliders due to an
unfortunate suppression of the coupling between the $Z$ boson and leptons. This motivates us to
look beyond the leading order.

In this paper, we thus investigate $W$ boson fusion at $\mu^+ \mu^+$ colliders at higher order in perturbation theory\footnote{
Here, we mean ``higher order'' with respect to the leading-order $W$ boson fusion process at $\mu^+ \mu^-$ colliders, i.e.,
$\mu^+ \mu^- \to \overline{\nu}_{\mu} \nu_{\mu} h$, rather than higher-order corrections to some process at $\mu^+ \mu^+$ colliders.
}.
The pertinent process, which we show in Fig.~\ref{fig:wbf}, begins with emission of a photon or $Z$ boson from
one of the antimuons, followed by the photon or $Z$ boson splitting into a $W^+ W^-$ pair. The $W^-$ from this splitting then collides with the $W^+$ emitted from the other antimuon to produce a Higgs boson.
For large center-of-mass energies $\sqrt{s}$, we find that this higher-order process involves a $(\log s)^3$ factor, in contrast to the single $\log s$ appearing in the leading-order $W$ boson fusion at $\mu^+ \mu^-$ and $e^+ e^-$ colliders~\cite{AlAli:2021let, ILC:2013jhg}. Due to this large enhancement, the cross section for Higgs boson production at $\mu^+ \mu^+$ colliders becomes comparable to that at $\mu^+ \mu^-$ colliders for $\mathcal{O}(10)$~TeV beam energies.

To compute the part of the process involving photon emission, we must carefully address infrared divergences, which are physically cut off by the muon mass.
However, directly using numerical codes such as the event generator
MadGraph~\cite{Alwall:2014hca} leads to instabilities
in the numerical phase-space integration, since the muon mass is either set to zero,
or significantly smaller than the center-of-mass energy.
Therefore, we also discuss how to reliably and accurately compute fixed-order cross sections for such infrared-divergent processes.

We furthermore discuss the size of subsequent higher-order processes
to determine if the fixed-order computation is meaningful. We find
that at $10$~TeV energies, the processes with further emissions of extra
gauge bosons are much smaller than the process of our interest, such that fixed-order calculations are still valid.

Lastly, it is important for actual experiments whether the final-state $W^+$ has a transverse momentum large enough to be visible. To assess this, we generate sample
events using MadGraph and find that for $\sqrt s = 2$~TeV or $10$~TeV,
the vast majority of jets from the $W^+$ decays can be detected. This
means that it is possible to analyze the $W$ boson fusion process with a
$W^+ h$ final state separately from $Z$ boson fusion to perform
precision measurements of the Higgs boson couplings.

At the same time, the large size of the $\gamma$- and $Z$-mediated process
means that an analogous process gives a large contribution to the Higgs
production cross section also at $\mu^+ \mu^-$ and $e^+e^-$ colliders.
One should thus include this higher-order process in the
simulation and analysis of the Higgs boson production for coupling
measurements, as it may contaminate the events of the pure $W$ boson fusion
process.

This paper is organized as follows.
In Sec.~\ref{sec:wbf_mu+mu+}, we estimate the cross section of the $\gamma$- and $Z$-mediated Higgs boson production process at $\mu^+ \mu^+$ colliders
analytically at the leading-logarithm level, as well as numerically, and 
discuss its high-energy behavior.
We then explain a more reliable method to obtain the cross section in Sec.~\ref{sec:calc}, where we confirm the large enhancement of the cross section.
Next, since the $\gamma$- and $Z$-mediated process gives an extra $W$ boson in the final
state, we simulate collider events and discuss the possibility of identifying the process in Sec.~\ref{sec:event_shape}.
Finally, we summarize our findings in Sec.~\ref{sec:summary}.

In Appendix~\ref{sec:iww_error}, we review the derivation of the formula of the equivalent photon approximation, and discuss the uncertainties of the approximation.
We also discuss the error associated with the treatment of neglecting the muon mass
in the numerical calculation in Appendix~\ref{sec:mass_error}. 
Lastly, we detail a semi-automatic numerical method to implement the calculation explained
in Sec.~\ref{sec:calc}
in Appendix~\ref{sec:madgraph}.


\section{\texorpdfstring{$\gamma$}{gamma}- and \texorpdfstring{$Z$}{Z}-mediated \texorpdfstring{$W$}{W} boson fusion process at \texorpdfstring{$\mu^+ \mu^+$}{mu+mu+} colliders}
\label{sec:wbf_mu+mu+}

In this section, we discuss how $W$ boson fusion processes are possible at
$\mu^+ \mu^+$ colliders, and why they are important at high energies.
Therein, we derive an analytic formula for the cross section at the
leading-logarithm level. We also calculate the cross section
numerically using the event generator MadGraph~\cite{Alwall:2014hca} with a setting for parton
distribution functions of the photon in one of the
muons.
Note that both methods have uncertainties related to their scale settings, and the results 
should therefore be viewed as estimations. We will thus discuss a method to obtain
the cross section with controlled uncertainties in the
next section. Lastly, in this section, we compare the approximate results with those obtained
by the more reliable and accurate computation method, and show that the $\gamma$- and $Z$-mediated
$W$ boson fusion process is indeed important.


\subsection{Triple logarithms in the Higgs boson production}
\label{subsec:HPR-log_counting}

At high-energy lepton colliders, a large amount of Higgs bosons can be
produced through vector boson fusion.
When two antimuons collide with center-of-mass energy $\sqrt{s}
\gtrsim 1$~TeV, $Z$ boson fusion is enhanced by a factor of $\log s$,
similarly to $Z$ and $W$ boson fusion processes at $\mu^+\mu^-$ and $e^+e^-$ colliders.

By contrast, there is no coupling between $\mu^+$ and $W^-$ in the Standard Model
at the first order in the electromagnetic coupling $e$ or weak coupling $g$. This makes $W$ boson fusion difficult, since it requires a $W^+ W^-$ pair.
In particular, while the $W$ boson fusion process may take place at higher orders in
perturbation theory, these are seemingly suppressed by additional powers of the small couplings.
Therefore, we na\"{i}vely do not expect a satisfyingly large contribution to Higgs production from
$W$ boson fusion at $\mu^+ \mu^+$ colliders.
However, this expectation turns out to be incorrect, as we will see below.

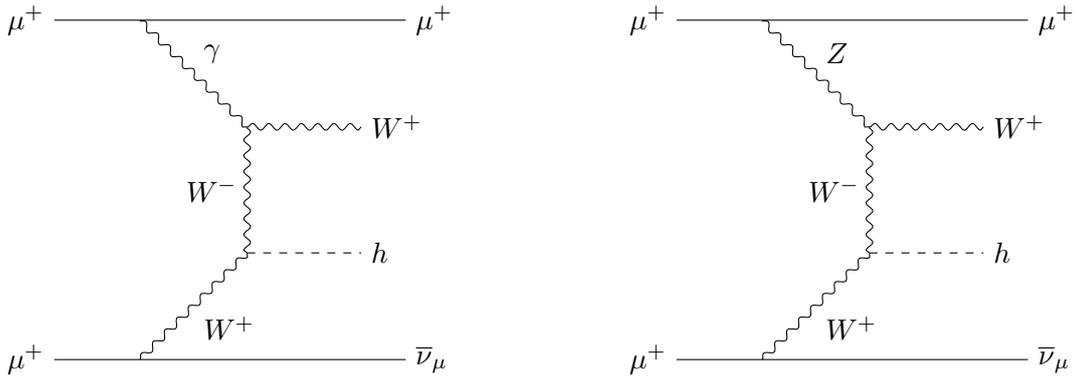
\begin{figure}[t!]
    \centering
    \begin{minipage}[b]{0.45\textwidth}
        \centering
        \begin{tikzpicture}
            \begin{feynman}
                \vertex (mu1) {\(\mu^{+}\)};
                \vertex [right=1.5cm of mu1] (intMu1);
                \vertex [right=3.5cm of intMu1] (mu1p) {\(\mu^+\)};
                \vertex [below=4.5cm of mu1] (mu2) {\(\mu^{+}\)};
                \vertex [right=1.5cm of mu2] (intMu2);
                \vertex [right=3.5cm of intMu2] (nu2) {\(\overline \nu_\mu\)};
                
                \vertex [below right=2cm of intMu1] (intAWW);
                \vertex [right=of intAWW] (w+) {\(W^+\)};
                \vertex [above right=2cm of intMu2] (intWWH);
                \vertex [right=of intWWH] (h) {\(h\)};
                
                \diagram* {
                (mu1) -- [plain] (intMu1) -- [plain] (mu1p),
                (mu2) -- [plain] (intMu2) -- [plain] (nu2),
                (intMu1) -- [boson, edge label=\(\gamma\)] (intAWW),
                (intAWW) -- [boson] (w+),
                (intAWW) -- [boson, edge label'=\(W^-\)] (intWWH),
                (intMu2) -- [boson, edge label'=\(W^+\)] (intWWH),
                (intWWH) -- [scalar] (h),
                };
            \end{feynman}
        \end{tikzpicture}
    \end{minipage}
    \hfill
    \begin{minipage}[b]{0.45\textwidth}
        \centering
        \begin{tikzpicture}
            \begin{feynman}
                \vertex (mu1) {\(\mu^{+}\)};
                \vertex [right=1.5cm of mu1] (intMu1);
                \vertex [right=3.5cm of intMu1] (mu1p) {\(\mu^+\)};
                \vertex [below=4.5cm of mu1] (mu2) {\(\mu^{+}\)};
                \vertex [right=1.5cm of mu2] (intMu2);
                \vertex [right=3.5cm of intMu2] (nu2) {\(\overline \nu_\mu\)};
                
                \vertex [below right=2cm of intMu1] (intAWW);
                \vertex [right=of intAWW] (w+) {\(W^+\)};
                \vertex [above right=2cm of intMu2] (intWWH);
                \vertex [right=of intWWH] (h) {\(h\)};
                
                \diagram* {
                (mu1) -- [plain] (intMu1) -- [plain] (mu1p),
                (mu2) -- [plain] (intMu2) -- [plain] (nu2),
                (intMu1) -- [boson, edge label=\(Z\)] (intAWW),
                (intAWW) -- [boson] (w+),
                (intAWW) -- [boson, edge label'=\(W^-\)] (intWWH),
                (intMu2) -- [boson, edge label'=\(W^+\)] (intWWH),
                (intWWH) -- [scalar] (h),
                };
            \end{feynman}
        \end{tikzpicture}
    \end{minipage}
    \caption{We show representative Feynman diagrams of single Higgs boson production
    through quasi-real photon (left) or $Z$ boson (right) emission, splitting, and subsequent $W$ boson fusion.}
    \label{fig:wbf}
\end{figure}

We show two representative Feynman diagrams for Higgs boson
production via such a higher-order $W$ boson fusion in Fig.~\ref{fig:wbf}.
Here, one antimuon provides a $W^+$, and the other a $W^-$ through pair
production by an intermediate photon or $Z$ boson.
When beam energies are much larger than the electroweak scale given by the vacuum expectation
value of the Higgs field, $v \simeq 246$~GeV, these
two diagrams are identical up to terms of the order of $m_Z^2 / s$, where $m_Z$
is the mass of the $Z$ boson. Note that one needs to, however, account for the different couplings
of the photon and $Z$ boson, as well as the infrared scales of their respective emission.

First, we focus on the contribution of the photon, ignoring the $Z$ boson.
Quantum electrodynamics enables us to factorize
the contribution of the photon into a parton distribution function with respect to the antimuon beam,
and a partonic cross section with an on-shell photon in the initial state.
The corresponding parton distribution function is given by
\begin{equation}
    f_{\gamma/\mu}(x, \mu_f^2) = \frac{\alpha}{2 \pi} \frac{1 + (1 - x)^2}{x}
     \log \frac{\mu_f^2}{m_\mu^2}
    \label{eq:a_mu}
\end{equation}
at the leading-logarithm
level~\cite{Weizsacker:1934,Williams:1935dka}, where $\alpha = e^2 / 4
\pi$ is the fine structure constant, $x$ is the longitudinal momentum fraction carried by the photon, $\mu_f$ is the factorization
scale, and $m_\mu$ the muon mass.
The large logarithm originates from an integral over the transverse momentum $p_{\rm T}$
of the differential parton distribution function
\begin{equation}
    \frac{{\rm d}f_{\gamma/\mu}}{{\rm d}p_{\rm T}^2} 
    = \frac{\alpha}{2 \pi} \frac{1 + (1 - x)^2}{x} \frac{1}{p_{\rm T}^2}\, ,
\end{equation}
which comprises a $p_{\rm T}^{-4}$ from the propagator of the photon and a $p_{\rm T}^2$ from the splitting amplitude.
Here, we adopt the muon mass $m_\mu$ as infrared scale of the emission, which physically cuts off the associated collinear divergence.

Similarly, one can also derive the parton distribution function of the $W$ boson. For its longitudinal component, this yields~\cite{Dawson:1984gx,Kane:1984bb}
\begin{equation}
    f_{W^+_{\rm L}/\mu^+_R}(x) = \frac{\alpha}{2 \pi \sin^2 \theta_{\rm W}} \frac{1 - x}{x}
    \quad \text{and} \quad f_{W^+_{\rm L}/\mu^+_L}(x) = 0\, ,
    \label{eq:w_mu}
\end{equation}
where $\theta_{\rm W}$ is the weak mixing angle,
also known as Weinberg angle.
The Nambu--Goldstone bosons eaten by the longitudinally polarized $W$ bosons have a typical scale $v$,
which modifies the dependence of the differential parton distribution function on the transverse momentum
from $1 / p_{\rm T}^2$ to $\sim v^2 / p_{\rm T}^4$~\cite{Chen:2016wkt}.
The integral over $p_{\rm T}^2$ thus does not lead to a large logarithm in the parton distribution
function of Eq.~\eqref{eq:w_mu}.

To estimate the full cross section of the process as seen on the left in Fig.~\ref{fig:wbf} using these parton distribution functions, we also need the cross section of the subprocess $\gamma W^+ \to W^+ h$, 
which is given by
\begin{equation}
    \label{eq:cs_aw}
    \sigma_{\gamma W} (s) = \frac{\pi \alpha^2}{m_W^2 \sin^2 \theta_{\rm W}}
    + \mathcal{O} \left( \frac{1}{s} \right)
\end{equation}
in the high energy limit $s \gg s_{\rm min} \equiv (m_h + m_W)^2$~\cite{HAGIWARA1992187}. Here,
$m_W$ and $m_h$ are the masses of $W$ and Higgs bosons, respectively.
We obtain this cross section by averaging over the polarizations of the initial photon, and summing over the ones of the final-state $W^+$, while taking the initial-state $W^+$ to be longitudinally polarized.
This is because the high-energy cross section is dominated by the contribution of order $s^0$, which comes from
the longitudinally polarized $W$ boson in the initial state,
since its polarization vector is almost proportional to its momentum. By contrast, the contribution from a transversely polarized $W$ boson
falls off as $1/s$ at high energies. Therefore, it does not contribute in the high-energy limit, and we may assume a longitudinally polarized $W$ boson in the initial state.

Finally, the photon contribution to the process $\mu^+ \mu^+ \to \mu^+ \overline{\nu}_\mu W^+ h$ is thus approximately given by
\begin{equation}
    \begin{split}
        \sigma_\gamma (s) &\simeq 2 \int\limits_{xys > s_{\rm min}}
         \! {\rm d}x {\rm d}y \, f_{\gamma/\mu^+} (x, \mu_f^2)
        \, f_{W^+_{\rm L}/\mu^+} (y) \, \sigma_{\gamma W} (x y s) \\[5pt]
        &\sim \frac{(1 + P_{\mu^+}) \, \alpha^4}{4 \pi m_W^2 \sin^4 \theta_{\rm W}}
        \left[ \log \frac{s}{m_\mu^2} \left(\log \frac{s}{s_{\rm min}}\right)^2
        - \frac{2}{3} \left(\log \frac{s}{s_{\rm min}}\right)^3
        \right]
    \end{split}
    \label{eq:Leading_Log_Approximation}
\end{equation}
at the leading-logarithm level, where $P_{\mu^+}$ denotes the
polarization of the antimuon beams. The factor of two in the first line accounts for photon or $W^+$ emission by either antimuon. For this estimate, we take the
factorization scale to be the center-of-mass energy at the parton
level, $\mu_f^2 = x y s$.

In Eq.~\eqref{eq:Leading_Log_Approximation}, we observe three powers of logarithms increasing with the center-of-mass energy. These originate from collinear and soft divergences arising in the limit of massless muons, $W$ bosons, and Higgs bosons. In particular, the photon emission from one $\mu^+$ has an associated collinear divergence $\sim\log s/m_\mu^2$ and soft divergence $\sim\log s/s_{\rm min}$, stemming respectively from the logarithm and $1/x$ factor in the parton distribution function of Eq.~\eqref{eq:a_mu}. Similarly, the $W^+$ emitted by the other $\mu^+$ has an associated soft divergence $\sim\log s/s_{\rm min}$, but no collinear divergence since it is longitudinally polarized, and thus its parton distribution function in Eq.~\eqref{eq:w_mu} only has a $1/x$ factor and no collinear logarithm. The second term $\sim(\log s/s_{\rm min})^3$ in the brackets of Eq.~\eqref{eq:Leading_Log_Approximation} arises because our chosen factorization scale depends on momentum fractions, and would for instance not appear for $\mu_f^2 = s$. Our choice of factorization scale, and hence the additional term, serves to improve the (numerical) accuracy of the parton distribution function approximation.

We thus find three large logarithms in the final formula. Compared with the cross
section of the leading-order $Z$ boson fusion process~\cite{Cahn:1983ip,Cahn:1984tx},
\begin{equation}
    \sigma_{\rm ZBF} (s) \sim \frac{\alpha^3 (1 - 4 \sin^2 \theta_{\rm W} + 8 \sin^4 \theta_{\rm W})^2}{64 m_Z^2 \sin^6 \theta_{\rm W} \cos^6 \theta_{\rm W}}
    \log \frac{s}{m_h^2}\, ,
\end{equation}
we observe a rapid growth of the cross section as a function of the
collider energy.
Furthermore, the cross section of the $Z$ boson fusion process has an unfortunate
suppression factor in the numerator. Because of this, at $\sqrt s \sim
\mathcal{O}(10)$~TeV, the higher-order, $\gamma$-mediated $W$ boson fusion cross section becomes much
larger than for the leading-order $Z$ boson fusion process.
Even when compared to the cross section of the leading-order $W$ boson
fusion process at $\mu^+ \mu^-$ or $e^+ e^-$
colliders~\cite{Cahn:1983ip,Cahn:1984tx},
\begin{equation}
    \sigma_{\rm WBF} (s) \sim \frac{\alpha^3}{16 m_W^2 \sin^6 \theta_{\rm W}}
    \log \frac{s}{m_h^2}\, ,
    \label{eq:Leading_Order_W_Fusion}
\end{equation}
the $\gamma$- (and $Z$-) mediated process becomes important at high energy
as two extra logarithmic factors can compensate the $\mathcal{O}(\alpha / \pi)$
suppression. For reference, the logarithmic factors of Eq.~\eqref{eq:Leading_Log_Approximation} and Eq.~\eqref{eq:Leading_Order_W_Fusion} at $10$~TeV evaluate to approximately $1100$ and $10$, respectively.

This may resemble a breakdown of perturbation theory of fixed-order computations. We will therefore return to the discussion of further
higher-order processes later in this section.


\subsection{Contributions from the \texorpdfstring{$Z$}{Z}-mediated process} 

At high energies, the contribution of the $Z$ boson becomes important. However, since
the $\gamma$- and $Z$-mediated diagrams interfere, we cannot simply
add the $Z$-mediated cross section, estimated using a
parton distribution function of the $Z$ boson, to the $\gamma$-mediated one. Instead, a matrix form of parton distribution functions needs
to be used~\cite{Ciafaloni:2005fm, Chen:2016wkt, Bauer:2017isx}. Here, 
we therefore discuss how to estimate the additional contributions at the
leading-logarithm order.

If one could treat the $Z$ boson as a massless particle just like
photon, the amplitudes in Fig.~\ref{fig:wbf} should have the same form, i.e., 
\begin{equation}
    \mathcal{M} = \mathcal{M}_\gamma + \mathcal{M}_Z = g_\gamma \mathcal{M}_0 + g_Z \mathcal{M}_0
\end{equation}
for each chirality of the antimuons,
where $g_{\gamma}$ and $g_Z$ are the respective couplings of the photon and $Z$ boson to the antimuons.
The ratios of $\gamma$ and $Z$ couplings to antimuons and $W$ bosons are given by
\begin{equation}
    \frac{g^{\mu^+_R}_Z}{g^{\mu^+}_\gamma} = \frac{\frac{1}{2} - \sin^2 \theta_{\rm W}}{\sin \theta_{\rm W} \cos \theta_{\rm W}} \simeq 0.668\, ,
    \quad \frac{g^{\mu^+_L}_Z}{g^{\mu^+}_\gamma} = \frac{- \sin^2 \theta_{\rm W}}{\sin \theta_{\rm W} \cos \theta_{\rm W}} \simeq -0.535\, ,
\end{equation}
and
\begin{equation}
    \frac{g^W_Z}{g^W_\gamma} = \frac{1}{\tan \theta_{\rm W}} \simeq 1.87
\end{equation}
for $\sin^2 \theta_{\rm W} \simeq 0.222$, where we have taken the value used by MadGraph for consistency with later computations.
Since the full matrix element can be expressed in terms of a shared amplitude $\mathcal{M}_0$ when taking the massless-$Z$ limit,
we can approximate the squared matrix element as the sum of three squared amplitudes as
\begin{equation}
    \vert \mathcal{M} \vert^2 = \vert g_\gamma \vert^2 \vert \mathcal{M}_0 \vert^2
    + \vert g_Z \vert^2 \vert \mathcal{M}_0 \vert^2 + 2 {\rm Re} (g_\gamma g_Z) \vert \mathcal{M}_0 \vert^2 + \mathcal{O}(m_Z^2 / s)\, .
\end{equation}

This result allows us to estimate the cross section by using the three parton distribution functions of the photon, the $Z$ boson, and their mixing for the corresponding terms. These functions satisfy
\begin{equation}
    \frac{{\rm d}f_{Z_{\rm T}/\mu^+_R}}{{\rm d}p_{\rm T}^2}
    = \left( \frac{g^{\mu^+_R}_Z}{g^{\mu^+}_\gamma} \right)^{\!\!\! 2}
    \frac{{\rm d}f_{\gamma/\mu}}{{\rm d}p_{\rm T}^2},
    \quad \frac{{\rm d}f_{Z_{\rm T}/\mu^+_L}}{{\rm d}p_{\rm T}^2}
    = \left( \frac{g^{\mu^+_L}_Z}{g^{\mu^+}_\gamma} \right)^{\!\!\! 2}
    \frac{{\rm d}f_{\gamma/\mu}}{{\rm d}p_{\rm T}^2}\, ,
\end{equation}
and
\begin{equation}
    \frac{{\rm d}f_{[\gamma Z_{\rm T}]/\mu^+_R}}{{\rm d}p_{\rm T}^2}
    = \frac{g^{\mu^+_R}_Z}{g^{\mu^+}_\gamma} \frac{{\rm d}f_{\gamma/\mu}}{{\rm d}p_{\rm T}^2}\, ,
    \quad \frac{{\rm d}f_{[\gamma Z_{\rm T}]/\mu^+_L}}{{\rm d}p_{\rm T}^2}
    = \frac{g^{\mu^+_L}_Z}{g^{\mu^+}_\gamma} \frac{{\rm d}f_{\gamma/\mu}}{{\rm d}p_{\rm T}^2}\, .
\end{equation}
Here, we have neglected the longitudinal components of the $Z$ boson since its parton distribution function does not have logarithmic enhancement, analogously to the longitudinal $W$ boson. (See the first expression in Eq.~\eqref{eq:w_mu}.)
Since the $Z$ boson contribution starts from $p_{\rm T}^2 \sim m_Z^2$, we can
obtain the respective correction by integrating these differential equations with
$f_{Z_{\rm T}/\mu} = f_{[\gamma Z_{\rm T}]/\mu} = 0$ at $p_{\rm T}^2 =
m_Z^2$ as the boundary condition.

At the leading-logarithm order, we thus obtain
\begin{equation}
    \sigma (s) = \sigma_\gamma (s) + \sigma_{Z_{\rm T}} (s) + \sigma_{[\gamma Z_{\rm T}]} (s)\, ,
\end{equation}
where
\begin{equation}
    \begin{split}
        \sigma_{Z_{\rm T}} (s) &= \sigma_\gamma (s) \, g (s) \left[ \frac{1 + P_{\mu^+}}{2}
        \left( \frac{g^{\mu^+_R}_Z}{g^{\mu^+}_\gamma} \frac{g^W_Z}{g^W_\gamma} \right)^{\!\!\! 2}
        + \frac{1 - P_{\mu^+}}{2}
        \left( \frac{g^{\mu^+_L}_Z}{g^{\mu^+}_\gamma} \frac{g^W_Z}{g^W_\gamma} \right)^{\!\!\! 2}
        \, \right] \\[3pt]
        &\simeq \sigma_\gamma (s) \, g (s) \, \left[ 1.28 + 0.281 \, P_{\mu^+} \right]\, ,
    \end{split}
\end{equation}
and
\begin{equation}
    \begin{split}
        \sigma_{[\gamma Z_{\rm T}]} (s) &= 2\, \sigma_\gamma (s) \, g (s) \left[ \frac{1 + P_{\mu^+}}{2}
        \frac{g^{\mu^+_R}_Z}{g^{\mu^+}_\gamma} \frac{g^W_Z}{g^W_\gamma} + \frac{1 - P_{\mu^+}}{2}
        \frac{g^{\mu^+_L}_Z}{g^{\mu^+}_\gamma} \frac{g^W_Z}{g^W_\gamma} \right] \\[3pt]
        &\simeq \sigma_\gamma (s) \, g (s) \left[ 0.250 + 2.25 \, P_{\mu^+} \right]\, .
    \end{split}
\end{equation}
Here, we define the ratio of logarithmic factors originating from the $p_{\rm T}^2$ integrals as
\begin{equation}
    g(s) = \frac{\log (s / m_Z^2) - \frac{2}{3} \log (s / s_{\rm min})}
    {\log (s / m_\mu^2) - \frac{2}{3} \log (s / s_{\rm min})}\, ,
\end{equation}
which, e.g., gives overall factors of about 0.189, 0.238 and 0.268 for $\sqrt{s} = 2, \, 10$ and $30$~TeV, respectively.

Thus, by adding up the contributions involving the $Z$ boson, we finally find
\begin{equation}
\sigma (s)\, \simeq\, \sigma_\gamma (s)\, \cdot\, \left[ 1 + \left( 1.53 + 2.53 \, P_{\mu^+} \right)
     \, g (s) \right],
     \label{eq:z-contribution}
\end{equation}
which is significantly enhanced for positively polarized antimuons due
to positive interference effects.
Together with the enhancement in
Eq.~\eqref{eq:Leading_Log_Approximation}, positive polarization of
the $\mu^+$ beams can enhance the cross section by more than a factor
of two.
It is important to note that a polarized muon beam may be available, e.g., at $\mu^+ \mu^+$ colliders with
ultra-slow muon technology.
In the following, we present the results in both the unpolarized case with
$P_{\mu^+} = 0$, and the polarized case with $P_{\mu^+} = +0.8$ as suggested in~\cite{Hamada:2022mua}.


\subsection{Numerical calculation with MadGraph}

The large enhancement by $(\log s)^3$ motivates us to calculate the
cross section more accurately and reliably, which we discuss in
the next section.
However, we first numerically estimate the cross sections using the event generator MadGraph, and compare them to the results of the full calculation and the cross
sections of Higgs boson production via the leading $W$ and $Z$ boson fusion processes at
$\mu^+\mu^-$ and $e^+e^-$ colliders.

We note here in particular that the $\gamma$- and $Z$-mediated Higgs production process
also exists at $\mu^+\mu^-$ and $e^+e^-$ colliders with the same cross
section as for $\mu^+\mu^+$ colliders. Therefore, its large enhancement means that there are large corrections to the leading-order cross
section of $W$ boson fusion at high energies.

When estimating the cross section of the photon-mediated process in MadGraph, it is possible to
use a parton distribution function for the photon in one of the antimuons. 
We use one of these parton distribution function options, the Improved Weizs\"{a}cker--Williams
(IWW) setting~\cite{Frixione:1993yw}, for the computations presented here.
Note that we do not use the parton distribution function approximation for $W^+$ from the other
antimuon, and instead treat $\gamma \mu^+ \to \overline{\nu}_\mu W^+ h$
as the hard process to be convoluted with the IWW parton distribution function of the photon in
the antimuon.
We then take the contribution from the $Z$-mediated process into account
by subsequently multiplying the obtained result by the factor in Eq.~\eqref{eq:z-contribution}.

In calculations with parton distribution functions, it is necessary to set the scale for the hard
process, also referred to as factorization scale $\mu_f$. Among several options implemented in
MadGraph, we take two extremal choices: the default setting \verb|-1|, given by the transverse mass after $k_{\rm T}$ clustering to a $2 \to 2$ system; and the setting \verb|4|, given by the partonic center-of-mass energy as $\mu_f = \sqrt{xs}$~\cite{Hirschi:2015iia}, which is a commonly used choice for the factorization scale. 
Note that these are settings for the dynamical scale choice in MadGraph, which is then used for the factorization scale. Here, dynamical means that the scale is determined on an event-by-event basis, e.g., via the respective momentum fraction $x$ in $\mu_f = \sqrt{xs}$ of the event. One may also set a fixed factorization scale, but we do not use this here.

\begin{figure}[t!]
  \centering
  \includegraphics[width=0.95\textwidth]{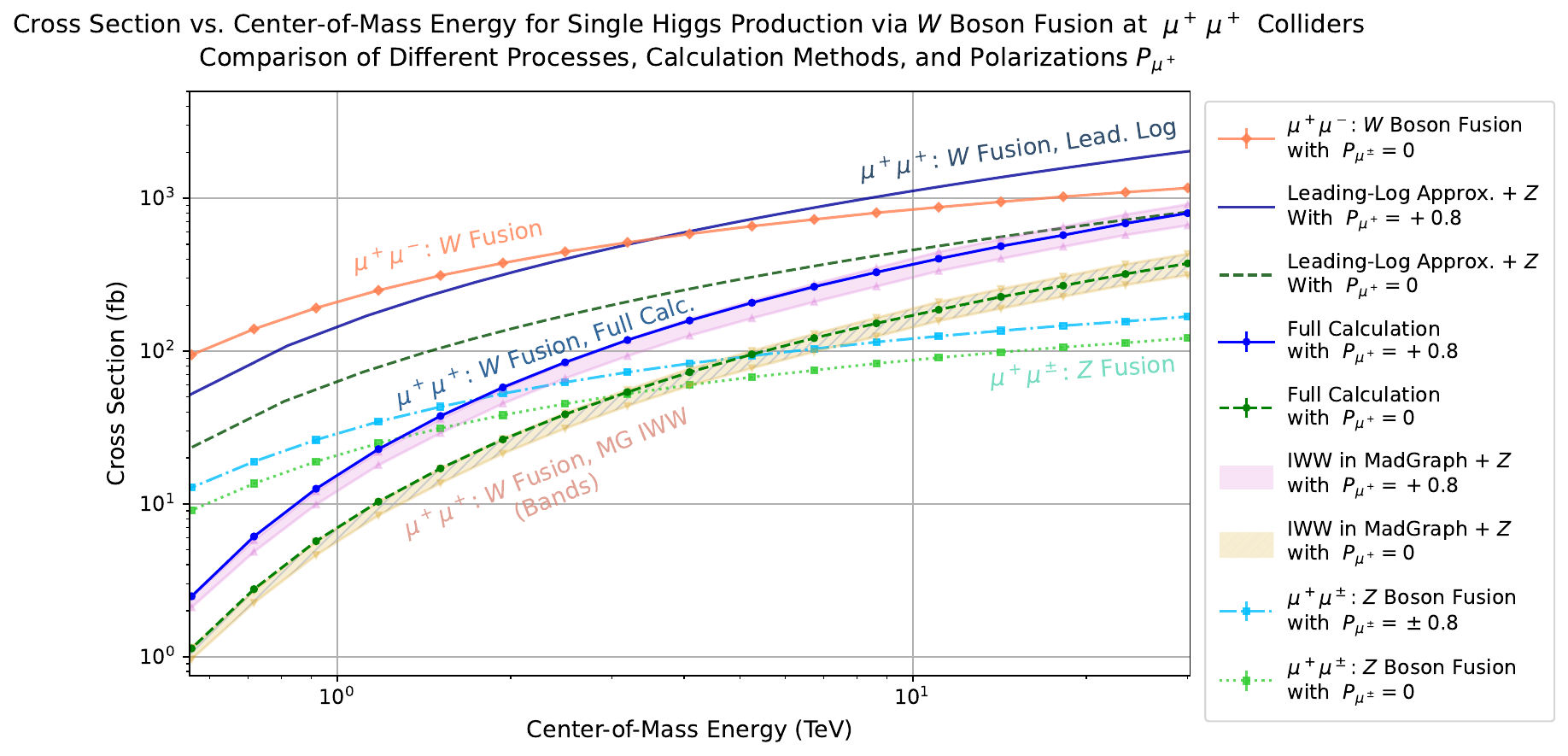}
  \cprotect\caption{We show the cross section for single Higgs boson production at $\mu^+ \mu^+$ colliders via $W$ boson fusion as a function of center-of-mass energy. We compare the results for this process obtained using different calculation methods and polarizations $P_{\mu^+}$ for both beams, as well as different processes for single Higgs production. First, we show the $W$ boson fusion at $\mu^+ \mu^-$ colliders, via $\mu^+ \mu^- \to
  \overline{\nu}_\mu \nu_\mu h$, with unpolarized beams, i.e., $P_{\mu^\pm} = 0$ (orange, solid with diamonds). Second, the leading-logarithm approximation of Eq.~\eqref{eq:Leading_Log_Approximation} (including the $Z$-boson factor of Eq.~\eqref{eq:z-contribution}) for $W$ boson fusion at
  $\mu^+ \mu^+$ colliders, via $\mu^+ \mu^+ \to \mu^+
  \overline{\nu}_\mu W^+ h$, with both beams +0.8 polarized, i.e., $P_{\mu^+} = +0.8$ (dark blue, solid), and
  unpolarized (dark green, dashed). Third, we show the full result for $W$ boson fusion at
  $\mu^+ \mu^+$ colliders with polarized (blue, solid with dots), and unpolarized (green, dashed with dots) beams. Fourth, the bands for the Improved Weizs\"acker--Williams (IWW) calculation in MadGraph for $W$ boson fusion at $\mu^+ \mu^+$ colliders, multiplied by the $Z$-boson factor of Eq.~\eqref{eq:z-contribution}, with polarized (purple, solid with upwards-pointing triangles) and unpolarized (yellow-green, dashed with downwards-pointing triangles) beams. The bounds of these bands are obtained by setting the dynamical scale in MadGraph to the standard setting \verb|-1|, which yields the lower bounds, and to the partonic center-of-mass energy denoted by the setting \verb|4|, giving the upper bounds. Lastly, we also show the $Z$ boson fusion process in $\mu^+ \mu^\pm$ collisions, via $\mu^+ \mu^\pm
  \to \mu^+ \mu^\pm h$, with polarized (light blue, dash-dotted with squares) and unpolarized (light green, dotted with squares) beams.}
  \label{fig:HPR-cross_section_results}
\end{figure}

In Fig.~\ref{fig:HPR-cross_section_results}, we show the cross section
of the $\gamma$- and $Z$-mediated Higgs production process, given by $\mu^+
\mu^+ \to \mu^+ \overline{\nu}_\mu W^+ h$.
For the estimate using MadGraph, the upper band represents the results with $P_{\mu^+} = +0.8$, while the lower one corresponds to $P_{\mu^+} = 0$.
We obtain the upper and lower ends of each band with
different choices of factorization scales for the hard process with an initial-state photon, as mentioned above. We then include the
contributions from $Z$ boson exchanges by multiplying these results by the
factor in Eq.~\eqref{eq:z-contribution} to obtain the bands shown in the figure.
The solid curves near the bands are the results from the accurate
calculation, which we explain in the next section.
We see that the calculation with the IWW parton distribution function
multiplied by the $Z$-boson factor gives reasonable estimates for the
cross section.

The leading-logarithm formula of
Eq.~\eqref{eq:Leading_Log_Approximation} with the $Z$-boson factor in
Eq.~\eqref{eq:z-contribution} is depicted as lines without data points (solid and dashed lines for $P_{\mu^+} = +0.8$ and $P_{\mu^+} = 0$, respectively), and
seems to overestimate the cross sections by a factor of two to three
even at $\sqrt s \sim 10$~TeV. 
This discrepancy stems largely from the sub-leading logarithmic terms
ignored in Eq.~\eqref{eq:Leading_Log_Approximation}. These originate partly from the low-$s$ behavior of the partonic cross section $\sigma_{\gamma W} (s)$ near the threshold, which effectively shifts $s_{\rm min}$ to a larger value; and partly from the terms dropped after
convolution of the partonic cross section with the parton distribution functions. 
However, we have confirmed that including the sub-leading terms brings down the cross section, and gives numerically consistent results compared with the ones obtained from MadGraph, even for $\sqrt s \sim \mathcal{O}(1)$~TeV. Therefore, while the sub-leading logarithms give noticeable corrections at $O(10)$~TeV energies, we can thus confirm the parametric $(\log s)^3$ behavior.

\begin{table}[t!]
    {
    \newcolumntype{C}[1]{>{\centering\arraybackslash}p{#1}}
    \renewcommand{\arraystretch}{1.5}
    \centering
    \begin{tabular}{l|ccccc}
        {Center-of-Mass Energy [TeV]} & {$1$} & {$2$} & {$3$} & {$10$} & {$30$} \\
        \hline
        $\sigma(\mu^+ \mu^- \to \overline{\nu} \nu h)$ \hspace{2.1em} [fb], $P_{\mu^\pm} = 0$ & $211$ & $385$ & $498$ & $842$ & $1165$ \\
        $\sigma(\mu^+ \mu^+ \to \mu^+ \overline{\nu} W^+ h)$ [fb], $P_{\mu^+} = +0.8$& $15.6$ & $61.0$ & $109$ & $371$ & $799$ \\
        $\sigma(\mu^+ \mu^+ \to \mu^+ \overline{\nu} W^+ h)$ [fb], $P_{\mu^+} = 0$& $7.05$ & $27.7$ & $49.9$ & $172$ & $374$ \\
        $\sigma(\mu^+ \mu^+ \to \mu^+ \mu^+ h)$ \hspace{0.7em} [fb], $P_{\mu^+} = +0.8$ & $29.0$ & $53.8$ & $70.4$ & $121$ & $168$ \\
        $\sigma(\mu^+ \mu^+ \to \mu^+ \mu^+ h)$ \hspace{0.7em} [fb], $P_{\mu^+} = 0$ & $20.9$ & $39.0$ & $50.9$ & $87.6$ & $122$ \\
    \end{tabular}
    \caption{We show the total cross section in the unit of fb for single Higgs production via $W$ boson fusion in $\mu^+ \mu^-$ collisions ($\mu^+ \mu^- \to \overline{\nu}_\mu \nu_\mu h$) with unpolarized beams, $W$ boson fusion in $\mu^+ \mu^+$ collisions ($\mu^+ \mu^+ \to \mu^+ \overline{\nu}_\mu W^+ h$) with both beams +0.8 polarized or unpolarized, and via $Z$ boson fusion in $\mu^+ \mu^+$ collisions ($\mu^+ \mu^+ \to \mu^+ \mu^+ h$) with both beams +0.8 polarized or unpolarized for different center-of-mass energies.}
    \label{tab:cross_section_results_comparison}
    }
\end{table}

We furthermore overlaid the cross section of the leading-order $W$ boson fusion process at
$\mu^+ \mu^-$ colliders, where we do not assume beam polarizations.
We see that at energies of $10$~TeV or above, the cross section of
the higher-order process with polarized beams becomes about half as large as the
leading one at $\mu^+ \mu^-$ colliders.
Therefore, the disadvantage due to missing $W$ boson fusion at the leading order is remedied by the higher-order process at $\sim 10$~TeV
energies.
Importantly, this large contribution to Higgs production is also present for $\mu^+
\mu^-$ and $e^+ e^-$ colliders. This will thus add a new significant process
at any such lepton colliders.

Lastly, we show in Fig.~\ref{fig:HPR-cross_section_results} the cross section of $Z$ boson fusion at $\mu^+ \mu^+$ colliders for
$P_{\mu^+} = +0.8$ and $P_{\mu^+} = 0$. We see that the cross section of the higher-order $W$ boson fusion process becomes larger than that of the leading $Z$ boson fusion
beyond a few TeV.

For comparison, we list the
cross sections for representative collider energies in Tab.~\ref{tab:cross_section_results_comparison}.
We computed these values using the same methods as the respective curves in Fig.~\ref{fig:HPR-cross_section_results}, i.e., using MadGraph for $\sigma(\mu^+ \mu^- \to \overline{\nu} \nu h)$ and $\sigma(\mu^+ \mu^+ \to \mu^+ \mu^+ h)$, and the full calculation with controlled theoretical uncertainties described in Sec.~\ref{sec:calc} for $\sigma(\mu^+ \mu^+ \to \mu^+ \overline{\nu} W^+ h)$. Note that in Tab.~\ref{tab:cross_section_results_comparison}, we omit the theoretical uncertainties of the cross sections, which are of $\mathcal{O}(1)$~fb or smaller.

The observed enhancement due to the large logarithms in $W$ boson fusion at a higher order may indicate a breakdown of the perturbative expansion. 
If this is the case, one needs to resum higher-order corrections stemming from multiple soft and collinear emissions. 
Such resummation effects are encoded in electroweak parton distribution functions  
by solving a set of coupled differential equations, called the electroweak DGLAP equations \cite{Ciafaloni:2001mu, Ciafaloni:2005fm, Chen:2016wkt, Bauer:2017isx, Bauer:2017bnh, Manohar:2018kfx, Fornal:2018znf, Han:2020uid, Han:2021kes, Azatov:2022itm, Garosi:2023bvq, Marzocca:2024fqb}.

Here, to verify the validity of our 
fixed-order calculation, we evaluate the cross section of the process $\mu^+ \mu^+ \to \mu^+ \mu^+ W^+ W^- h$. Parametrically, this process has a logarithmic enhancement factor of $\alpha^5\, (\log s)^5$, compared to $\alpha^4\, (\log s)^3$ for the process $\mu^+ \mu^+ \to \mu^+ \overline{\nu} W^+ h$. Thus, the expansion parameter is $\alpha\, (\log s)^2$, which is the largest correction factor we expect.
The size of the $\mu^+ \mu^+ \to \mu^+ \mu^+ W^+ W^- h$ cross section therefore serves as an indicator of the reliability
of the perturbative expansion.
Using the method we describe in
Sec.~\ref{sec:calc}, we obtain the cross section 
$\sigma \simeq 14$~fb at $\sqrt s = 20$~TeV for $P_{\mu^+}=+0.8$,
which is notably smaller than the cross section $\sigma \simeq
600$~fb of the lower-order process
$\mu^+ \mu^+ \to \mu^+ \overline{\nu} W^+ h$. 
Therefore, at least for Higgs boson production at $\mathcal{O}(10)$~TeV energies, such higher-order processes are not yet in the non-perturbative regime.
Note that we expect to lose perturbativity
at $\mathcal{O}(100)$~TeV energies, where the 
logarithmic enhancement factor combined with the 
coupling constant, $\sim(\alpha / \pi) \log (s/m_\mu^2) \log(s/s_{\rm min})$, becomes $\mathcal{O}(1)$.
Nevertheless, in the analysis of actual events, one should include the higher-order processes since they may contribute to the signals defined by lower-order ones.


\section{Calculation with controlled theoretical uncertainties}
\label{sec:calc}

We saw in the previous section that a large enhancement of the cross
section occurs at high energies.
We furthermore examined its leading-logarithm approximation and a numerical method implementing parton distribution functions;
however, it should be possible to calculate the cross section accurately
without uncertainties, as we are considering a fixed-order, tree-level
process.
One could try to calculate the full fixed-order process in MadGraph, but as previously mentioned, in that case one 
encounters numerical instabilities in the integration of the phase space for small
transverse momenta $p_{\rm T}$ of the antimuon emitting the photon---this is why we previously opted for using a parton distribution function instead.

Here, we therefore discuss a method to calculate the cross section accurately and reliably, and with
controlled theoretical uncertainties. The method is simply to split
the integral over the $p_{\rm T}$ of the final-state antimuon into two parts with $p_{\rm T} <
p_{\rm T}^{\rm (cut)}$ and $p_{\rm T} \geq p_{\rm T}^{\rm (cut)}$, respectively, for some value of $p_{\rm T}^{\rm (cut)}$. We calculate the first part
semi-analytically using the Weizs\"acker--Williams
method, also referred to as Equivalent Photon Approximation
(EPA)~\cite{Weizsacker:1934, Williams:1935dka, PhysRev.104.211,
Dalitz:1957dd, Brodsky:1970vk,
Arteaga-Romero:1971pai}, and the second part numerically using MadGraph, since the lower bound on $p_{\rm T}$ mitigates the numerical instabilities.
This means we calculate
\begin{eqnarray}
    \sigma &=& \sigma \big|_{p_{\rm T} < p_{\rm T}^{\rm (cut)}}
    + \sigma \big|_{p_{\rm T} \geq p_{\rm T}^{\rm (cut)}} \label{eq:CS-Overview_Split_Cross_PTC} \\[3pt]
    &\simeq& \sigma_{\rm EPA} \Big|_{p_{\rm T} < p_{\rm T}^{\rm (cut)}}
    + \sigma_{\rm MG} \Big|_{p_{\rm T} \geq p_{\rm T}^{\rm (cut)}}\,,
    \label{eq:CS-Overview_Split_Cross_IWW_MG}
\end{eqnarray}
where in the second line, we approximate the first
term by ignoring higher-order terms in $p_{\rm T}^{\rm (cut)}$, and the second term, by 
ignoring the muon mass.
While
splitting the phase-space integral itself is completely independent of
$p_{\rm T}^{\rm (cut)}$ no matter its value, the premise of EPA and MadGraph being
reliable in their respective regions does depend on the $p_{\rm T}^{\rm (cut)}$ at which we
split the phase space. 
As such, the accuracy and applicability of the different calculation methods
relies on choosing a suitable value for $p_{\rm T}^{\rm (cut)}$. Therefore, when calculating the split total
cross section using this method, we need to check whether the sum of the two cross sections of Eq.~\eqref{eq:CS-Overview_Split_Cross_IWW_MG} exhibits a plateau---i.e., a constant region---
as a function of $p_{\rm T}^{\rm (cut)}$.
This is the aforementioned insensitivity of Eq.~\eqref{eq:CS-Overview_Split_Cross_PTC}
with respect to the value of $p_{\rm T}^{\rm (cut)}$. If we
find such a plateau, we know on one hand that the combined calculation
scheme works, and on the other hand in which range of $p_{\rm T}^{\rm (cut)}$ it is
applicable. As we will see in the case of Higgs production at
$\mu^+ \mu^+$ colliders, this range is physically well-motivated, which makes
the application to other processes straightforward.

We note that splitting the phase space and calculating its parts with
a parton distribution function in one part and the full matrix element in another part is not an entirely
new approach. It has for instance been used in Ref.~\cite{Ruiz:2021tdt}
under the name of matrix-element matching to show the validity of
the parton distribution functions of transversely polarized weak bosons.


\subsection{Low-\texorpdfstring{$p_{\rm T}$}{pt} cross sections from EPA}
\label{subsec:CS-Low_pT_IWW}

\begin{figure}[t!]
  \centering
  \includegraphics[width=0.95\textwidth]{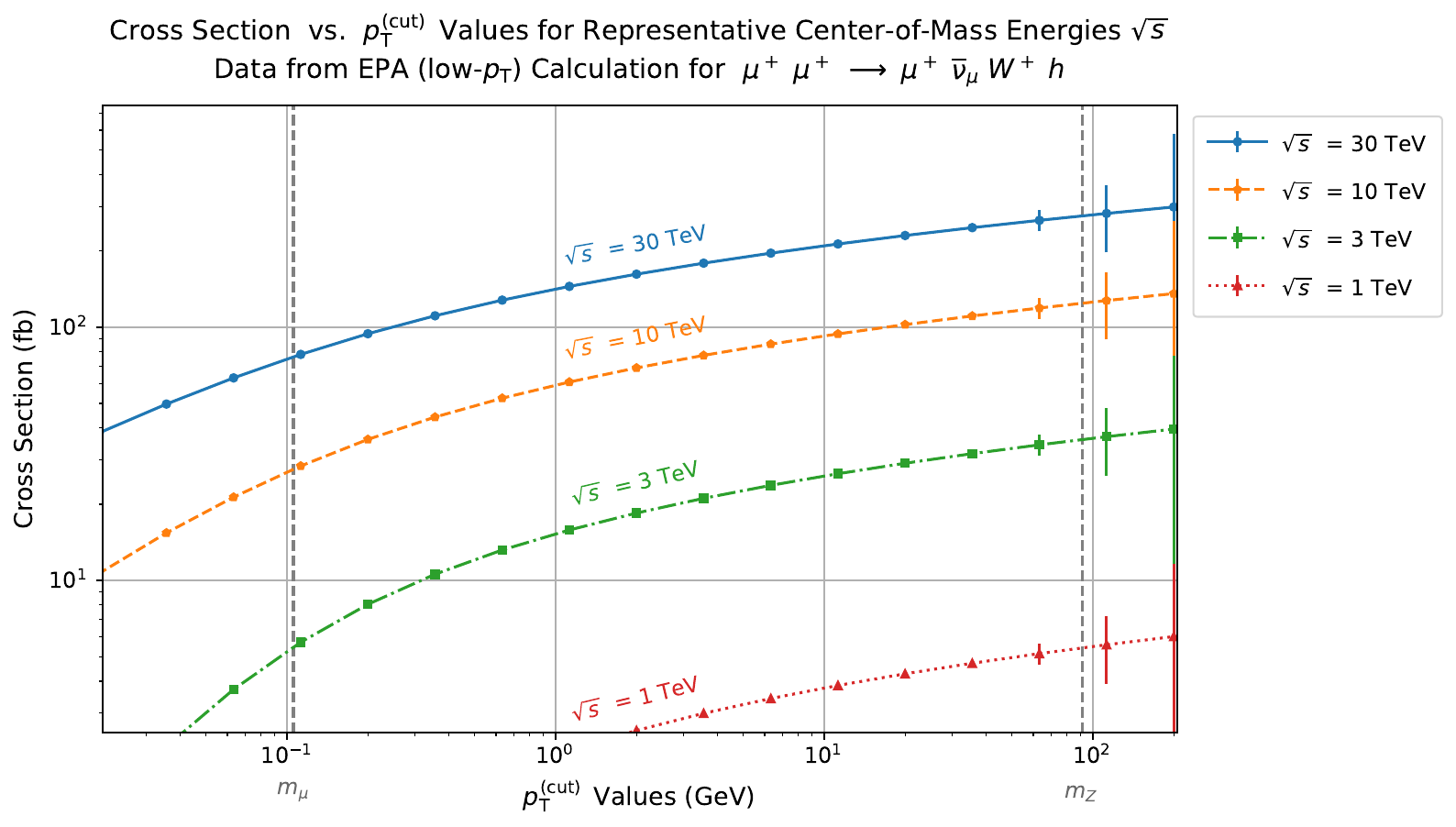}
  \caption{We show the EPA (low-$p_{\rm T}$) cross section for $\mu^+
  \mu^+ \to \mu^+ {\overline \nu_\mu} W^+ h$ as a function of $p_{\rm T}^{\rm (cut)}$ for representative center-of-mass energies (top to bottom): 30~TeV (blue, solid), 10~TeV (orange, dashed), 3~TeV (green, dash-dotted), and 1~TeV (red, dotted). The vertical dashed lines indicate the
  muon mass $m_\mu$ and the $Z$ boson mass $m_Z$. We calculate the error bars from the cross section value and the relative error of Eq.~\eqref{eq:iww_error}, combined with the supplied error of the numerical convolution, and the error supplied by MadGraph for the calculation of the partonic cross section after propagating it through the convolution.}
  \label{fig:HPR-IWW_Data_vs_pT}
\end{figure}

In our EPA calculation, we use the distribution function of the photon in the lepton $\ell$,
given by
\begin{eqnarray}
    f^{p_{\rm T}^{\rm (cut)}}_{\gamma/\ell} \!\!\! (x)
    &=& \frac{\alpha}{2 \pi} \left[ - \frac{2(1-x)}{x}
    \frac{\left(p_{\rm T}^{\rm (cut)}\right)^2}{\left(p_{\rm T}^{\rm (cut)}\right)^2 + m_\ell^2 x^2}
    + \frac{1 + (1-x)^2}{x} \log \frac{\left(p_{\rm T}^{\rm (cut)}\right)^2 + m_\ell^2 x^2}{m_\ell^2 x^2}\right]
    \nonumber \\
    &=& \frac{\alpha}{2 \pi} \left[ - \frac{2(1-x)}{x}
    + \frac{1 + (1-x)^2}{x} \log \frac{\left(p_{\rm T}^{\rm (cut)}\right)^2}{m_\ell^2\, x^2}\right]
    + \mathcal{O} \left( {m_\ell^2 \Big\slash \left( p_{\rm T}^{\rm (cut)} \right)^2} \right),
    \label{eq:iww}
\end{eqnarray}
where we put an upper bound on the transverse momentum via
$p_{\rm T}^{\rm (cut)}$, and $m_\ell$ is the mass of the lepton $\ell$. See Appendix~\ref{sec:iww_error} for the derivation. The first
expression is obtained by rewriting Eq.~(20) in
Ref.~\cite{Frixione:1993yw}, which is originally derived for
kinematical regions with small virtuality $q^2$ of the photon. We rewrite it using the relation of the photon virtuality and the transverse momentum, given by
\begin{equation}
    p_{\rm T}^2 = - q^2 (1-x) - m_\ell^2 x^2\, , 
\end{equation}
as we also discuss in Appendix~\ref{sec:iww_error}.
The second line of Eq.~\eqref{eq:iww}
is a further-approximated formula obtained when $p_{\rm T}^{\rm (cut)}$ is
much larger than the lepton mass $m_\ell$. This simplified formula can be
found in Ref.~\cite{Arteaga-Romero:1971pai}, again as a function of the
photon virtuality.
The first expression of the distribution function consistently takes
into account the finite lepton mass, up to
corrections of $\mathcal{O}(m_\ell^2/s)$. Note that although we use the first
expression of Eq.~\eqref{eq:iww} for our numerical calculations,
the second expression should actually be enough for our purposes by
taking $p_{\rm T}^{\rm (cut)} \gg m_\ell$. An important point is that we have a
non-logarithmic term, unlike the leading-logarithm formula in Eq.~\eqref{eq:a_mu},
for which there is no uncertainty associated with the scale of the hard process as long as the lepton mass is negligible---i.e., the second line of Eq.~\eqref{eq:iww}.

Using this parton distribution function, the total cross section for the low-$p_{\rm T}$ region is given by
\begin{equation}
    \sigma_{\rm EPA} \big|_{p_{\rm T} < p_{\rm T}^{\rm (cut)}}
    = \int_{s_{\rm min}/s}^1 {\rm d}x \, f^{p_{\rm T}^{\rm (cut)}}_{\gamma/\mu} \!\!\! (x)
    \, \sigma_{\gamma\mu}(xs)\, ,
    \label{eq:CS-IWW_pT_PDF_Derivation_0}
\end{equation}
where $\sigma_{\gamma\mu} (xs)$ is the cross section of the $\gamma \mu^+ \to
\overline{\nu}_\mu W^+ h$ process at the center-of-mass energy $\sqrt{xs}$.
The lower end of the $x$ integration is given by $s_{\rm min}/s$,
where $s_{\rm min}$ is the smallest invariant mass of the final-state
particles in the subprocess, i.e., $s_{\rm min} = (m_h + m_W)^2$, since the subprocess' center-of-mass energy needs to be large enough for it to take place, with $\sqrt{xs} \geq \sqrt{s_{\rm min}}$. The
cross section obtained from this formula is the leading-order value of
the systematic expansion in terms of $p_{\rm T}^{\rm (cut)}$. The relative uncertainty of this
method is of the order of
\begin{equation}
    \label{eq:iww_error}
    \frac{\left(p_{\rm T}^{\rm (cut)}\right)^2}{s_{\rm min}}\, .
\end{equation}
Note that EPA here is not the leading-logarithm approximation. The
uncertainty above can be much smaller than $\big[ \log \, \big(p_{\rm T}^{\rm (cut)} / m_\mu\big) \big]^{-1}$,
and the formula is valid for small $p_{\rm T}^{\rm (cut)}$ as long as $p_{\rm T}^{\rm (cut)} \gg m_\mu$.
We explain the derivation of the formula and the associated uncertainties in Appendix~\ref{sec:iww_error}.

The partonic cross section $\sigma_{\gamma \mu}$  can be calculated using MadGraph since there are no
intermediate light particles in this subprocess, and it can thus be evaluated numerically.
Therefore, we scan over center-of-mass energies to obtain the cross section as a function of energy, and then
convolute it with the parton distribution function, as described above.
For a small enough $p_{\rm T}^{\rm (cut)}$ compared to $s_{\rm min}$ (and large enough
compared to $m_\mu$), the cross section can thus be calculated with small
theoretical uncertainties. 
However, since we do not take into account the $Z$-boson-exchange diagrams using this method, we
expect it to be reliable for $p_{\rm T}^{\rm (cut)} \ll m_Z$.

We show in Fig.~\ref{fig:HPR-IWW_Data_vs_pT} the low-$p_{\rm T}$ cross
section as a function of $p_{\rm T}^{\rm (cut)}$ for representative center-of-mass energies $\sqrt{s} = 1$, $3$, $10$, and $30$~TeV. The error bars in the figure are calculated primarily using the relative uncertainty of Eq.~\eqref{eq:iww_error}, which is invisible for small $p_{\rm T}^{\rm (cut)}$.


\subsection{High-\texorpdfstring{$p_{\rm T}$}{pt} cross sections from MadGraph}
\label{subsec:CS-High_pT_MG}

\begin{figure}[t!]
  \centering
  \includegraphics[width=0.95\textwidth]{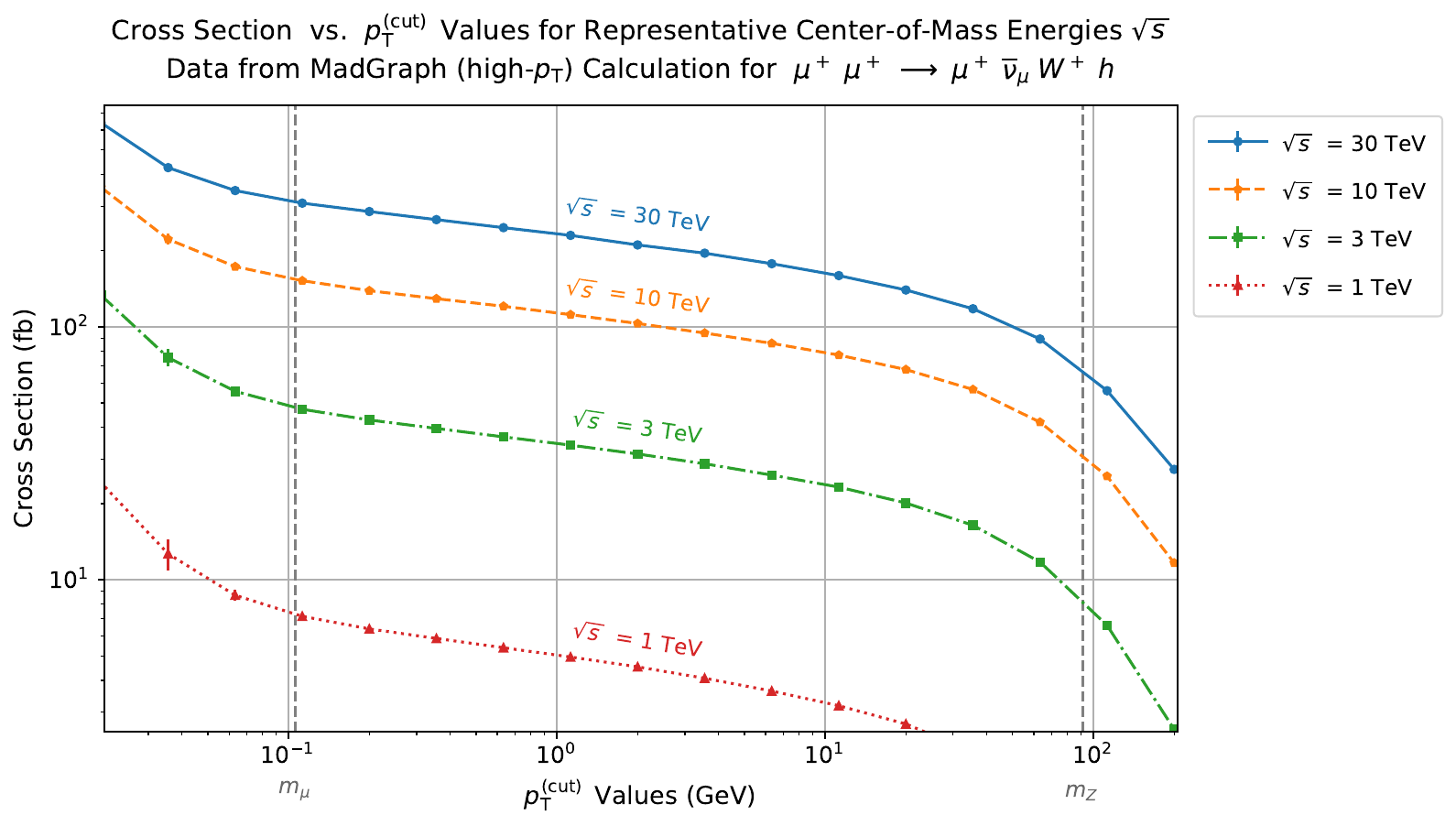}
  \caption{We show the MadGraph (high-$p_{\rm T}$) cross section for $\mu^+
  \mu^+ \to \mu^+ {\overline \nu_\mu} W^+ h$ as a function of $p_{\rm T}^{\rm (cut)}$ for representative center-of-mass energies (top to bottom): 30~TeV (blue, solid), 10~TeV (orange, dashed), 3~TeV (green, dash-dotted), and 1~TeV (red, dotted). The vertical lines indicate the
  muon mass $m_\mu$ and the $Z$ boson mass $m_Z$. We calculate the error bars from the cross section value and the relative error of Eq.~\eqref{eq:mg_error}, combined with the error supplied by MadGraph for the calculation of the cross section.}
  \label{fig:HPR-MadGraph_Data_vs_pT}
\end{figure}

The calculation for the high-$p_{\rm T}$ part can be  performed numerically
using MadGraph by specifying a lower bound on $p_{\rm T}$, given by $p_{\rm T}^{\rm (cut)}$. Thusly, we can avoid the collinear divergences (or numerical instabilities)
in the phase-space integral, which are associated with the muon mass being either set to zero, or significantly smaller than the center-of-mass energy.

In the default set-up, MadGraph calculates the cross section with the muon mass set to zero. This approximation
results in relative uncertainties of the order of
\begin{equation}
    \frac{m_\mu^2}{\left( p_{\rm T}^{\rm (cut)} \right)^2}\, 
    \left[ \log \frac{s}{\left( p_{\rm T}^{\rm (cut)} \right)^2}\,
    \log{\frac{s}{s_{\rm min}}} - \frac{2}{3} \left(\log{\frac{s}{s_{\rm min}}}\right)^2 \right]^{-1}.
    \label{eq:mg_error}
\end{equation}
See Appendix~\ref{sec:mass_error} for the derivation and discussion of the error associated with this massless approximation.
Therefore, by taking $p_{\rm T}^{\rm (cut)} \gg m_\mu$, we both avoid numerical
instabilities, and have negligibly small uncertainties associated with the finite muon
mass.

We show the obtained cross section as a function of $p_{\rm T}^{\rm (cut)}$ in Fig.~\ref{fig:HPR-MadGraph_Data_vs_pT} for the center-of-mass energies $\sqrt{s} = 1$, $3$, $10$, and $30$~TeV, analogously to the low-$p_{\rm T}$ cross section.


\subsection{Summing up the low- and high-\texorpdfstring{$p_{\rm T}$}{pt} cross sections}

\begin{figure}[t!]
  \centering
  \includegraphics[width=0.95\textwidth]{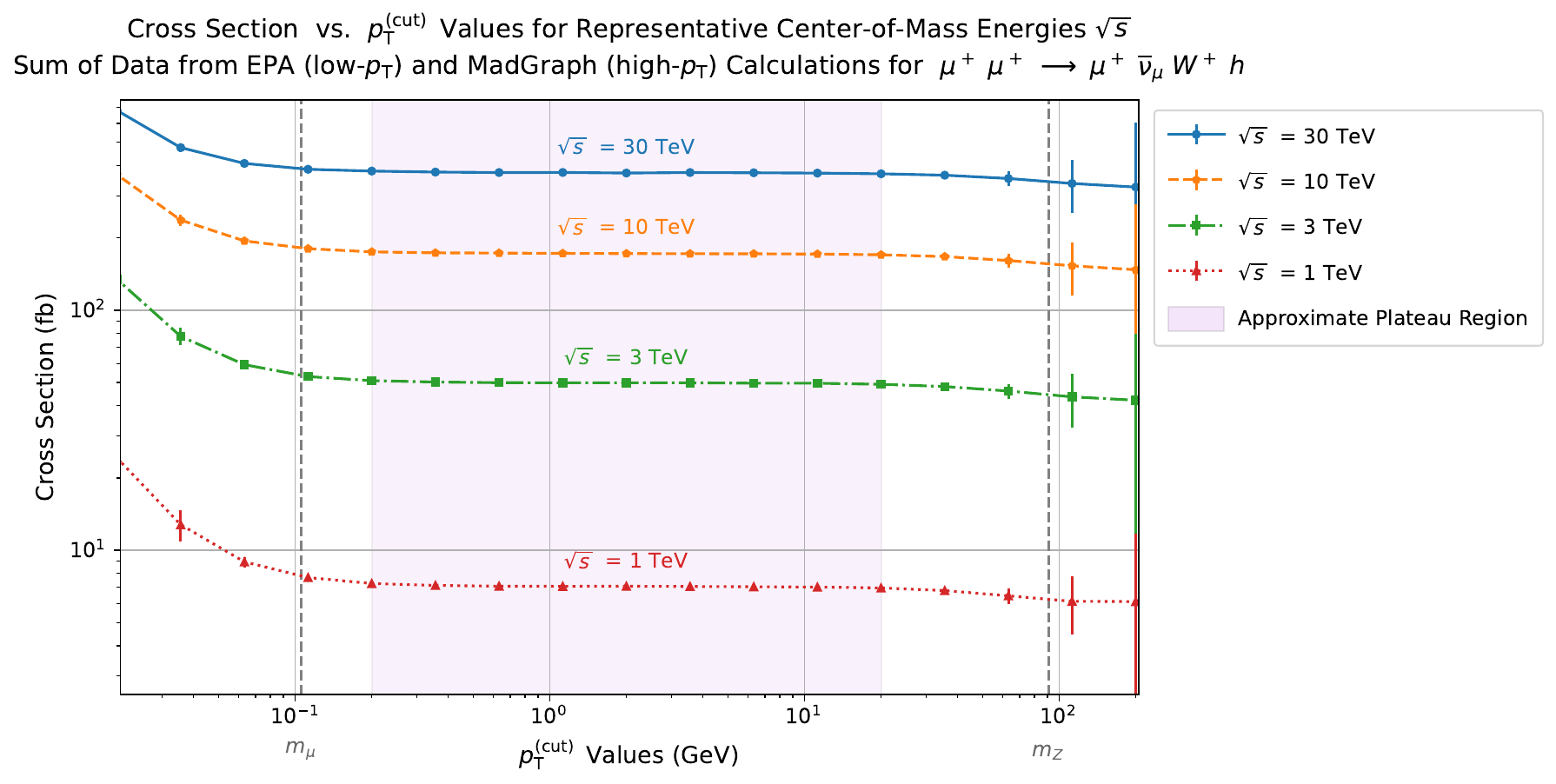}
  \caption{We show the sum of the EPA (low-$p_{\rm T}$) and MadGraph (high-$p_{\rm T}$) cross sections for $\mu^+
  \mu^+ \to \mu^+ {\overline \nu_\mu} W^+ h$ as a function of $p_{\rm T}^{\rm (cut)}$ for representative center-of-mass energies (top to bottom): 30~TeV (blue, solid), 10~TeV (orange, dashed), 3~TeV (green, dash-dotted), and 1~TeV (red, dotted). The vertical lines indicate the
  muon mass $m_\mu$ and the $Z$ boson mass $m_Z$. The error bars are determined by the root-square-sum of the EPA and MadGraph errors, which are primarily obtained from Eqs.~\eqref{eq:iww_error} and~\eqref{eq:mg_error}, respectively. The violet rectangle marks the approximate region of the plateau, i.e., the $p_{\rm T}^{\rm (cut)}$-independent region of the summed cross section, which we take as $0.2~{\rm GeV} \lesssim p_{\rm T}^{\rm (cut)} \lesssim 20~{\rm GeV}$.}
  \label{fig:HPR-Summed_Data_vs_pT}
\end{figure}

We show in Fig.~\ref{fig:HPR-Summed_Data_vs_pT} the sum of the low- and high-$p_{\rm T}$ cross
sections, calculated using the EPA scheme and MadGraph, respectively.
As expected, we clearly observe the plateau of the summed cross sections as a function of
$p_{\rm T}^{\rm (cut)}$.
Based on the discussions of the uncertainties in both calculations, we find an
optimal choice for $p_{\rm T}^{\rm (cut)}$ to be around the geometric mean of the muon and $Z$ boson masses, $p_{\rm T}^{\rm (cut)} = \sqrt{m_\mu m_Z} \simeq 3.10$~GeV,
where we indeed find the plateau.
Nevertheless, for too small or too large $p_{\rm T}^{\rm (cut)}$, the summed cross sections deviate from
their plateau values.

On the one hand, for small $p_{\rm T}^{\rm (cut)}$ (i.e., $\lesssim m_\mu$), the cross section calculated by MadGraph is too large due to numerical instabilities, as previously mentioned. 
However, the error bars in the small $p_{\rm T}^{\rm (cut)}$ region represent the uncertainties from the massless approximation, whereas the stark increase of the cross section calculated by MadGraph seems to be largely independent of this approximation.
Indeed, this behavior persists
even when including the finite muon mass in MadGraph. This is because the muon mass, while non-zero, is significantly smaller than the center-of-mass energies considered, leading again to numerical instabilities. 
Thus, even when including a finite muon mass, one cannot obtain a reliable result from MadGraph for, e.g., $p_{\rm T}^{\rm (cut)} = 0$.

On the other hand, for large $p_{\rm T}^{\rm (cut)}$ (i.e., $\gtrsim m_Z)$, the contribution from $Z$-boson-exchange diagrams,
as well as the higher-order terms in the $p_{\rm T}^{\rm (cut)}$ expansion of EPA become
important for the low-$p_{\rm T}$ cross section. Neglecting these thus leads to a deficit in the summed cross section.
Note that there are also diagrams without collinear
divergences, which are taken into account
in the MadGraph (high-$p_{\rm T}$) cross section. Increasing $p_{\rm T}^{\rm (cut)}$ too far thus also neglects part of their contribution.

Note that the results discussed here for unpolarized beams apply in the same way to the case with polarized beams.

\begin{figure}[t!]
  \centering
  \includegraphics[width=0.95\textwidth]{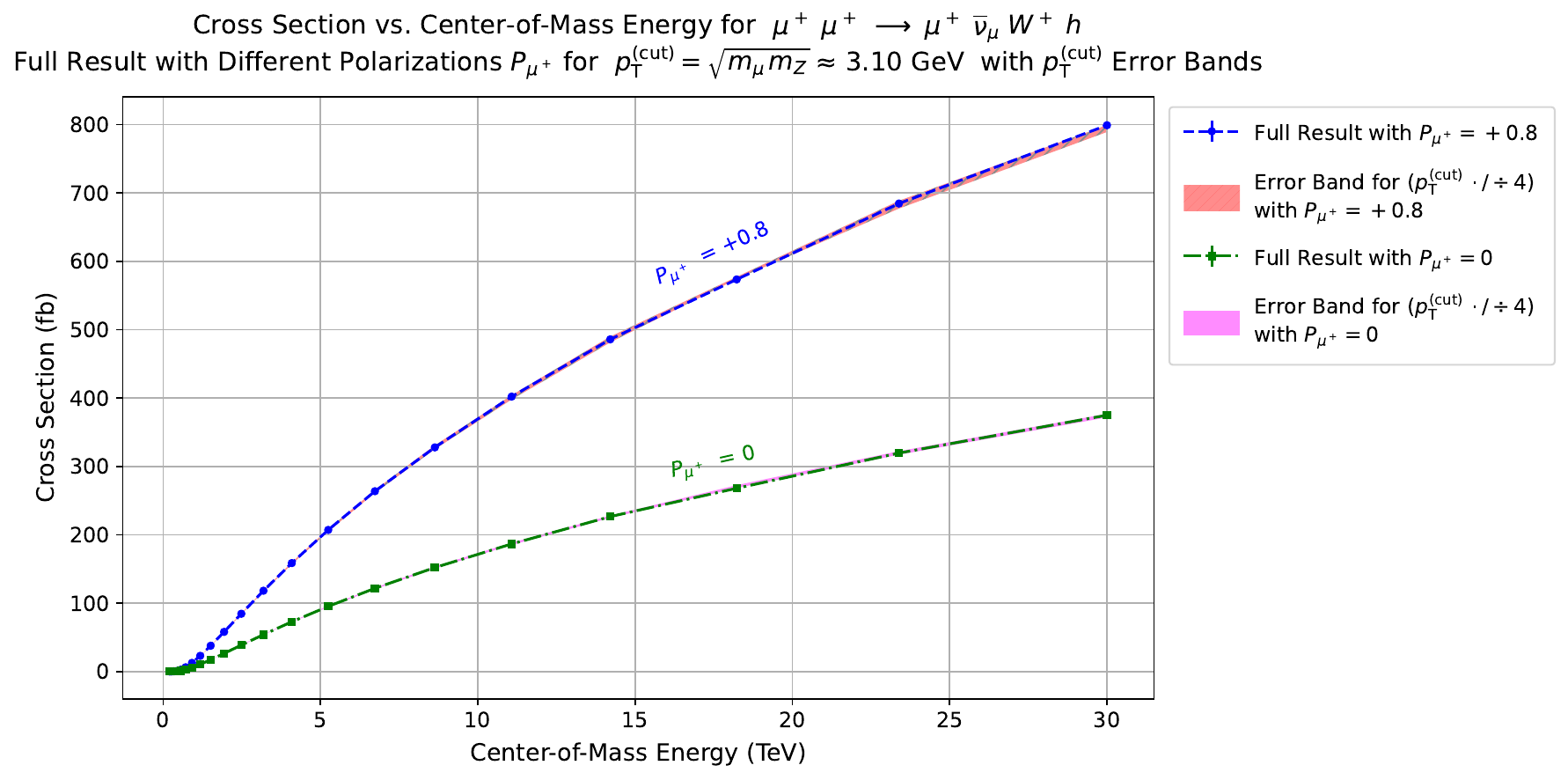}
  \caption{We show the full, summed cross section as a function of center-of-mass energy for $p_{\rm T}^{\rm (cut)} = \sqrt{m_\mu\, m_Z} \simeq 3.10$~GeV, the geometric mean of the muon mass $m_\mu$ and the $Z$ boson mass $m_Z$. We obtain the error bands by varying $p_{\rm T}^{\rm (cut)}$ in the range $\sqrt {m_\mu m_Z} / 4 < p_{\rm T}^{\rm (cut)} < 4 \sqrt {m_\mu m_Z}$. The upper curve (blue, dashed) and its error band (red, diagonal lines) correspond to a polarization of $P_{\mu^+} = +0.8$ for both beams, and the lower curve (green, dash-dotted) and its band (magenta, solid) to $P_{\mu^+} = 0$, i.e., unpolarized beams.}
  \label{fig:HPR-Summed_Data_vs_Energy}
\end{figure}

By reading off the cross sections in the plateau region, we can now
reliably calculate the fixed-order cross section of the $\mu^+ \mu^+
\to \mu^+ {\overline \nu_\mu} W^+ h$ process as
shown in Fig.~\ref{fig:HPR-Summed_Data_vs_Energy}.
The upper and lower curves are the cross sections with beam polarizations $P_{\mu^+}
= + 0.8$ and $P_{\mu^+} = 0$, respectively. We also show error bands associated with the choice of $p_{\rm T}^{\rm (cut)} = \sqrt{m_\mu m_Z}$ for the range $\sqrt{m_\mu
m_Z} / 4 < p_{\rm T}^{\rm (cut)} < 4 \sqrt{m_\mu m_Z}$. However, they are
almost invisible.
We thereby estimate the uncertainty associated with the choice of $p_{\rm T}^{\rm (cut)}$ to be less than one percent.

For convenience, we present in Appendix~\ref{sec:madgraph} a semi-automatic method to implement
the $p_{\rm T}^{\rm (cut)}$ scheme in MadGraph, in particular the low-$p_{\rm T}$ calculation.


\section{Event shape of the \texorpdfstring{${\bm W}$}{W} boson fusion process}
\label{sec:event_shape}

As seen so far, we obtained reliable results for the higher-order Higgs production
by dividing the cross section into two regions, corresponding to
low and high $p_{\rm T}$ of the final-state antimuon, which we calculate 
using EPA and MadGraph, respectively.
Compared to the leading-order $W$ boson fusion,
this process contains an additional $W$ boson in the final state.
Thus, to distinguish event signals from background, the produced Higgs and $W$ bosons should both be detected.

However, because the $\mu^+$ beams induce undesired backgrounds such as $\mu^+$-decay products and incoherent $e^+e^-$ pairs
(see Ref.~\cite{Casarsa:2023vqx} for a review),
we need to put shielding nozzles around the interaction point to reduce these beam-induced backgrounds.
This makes placement of detectors quite non-trivial, 
leading to constraints of the detectors' coverage angle. 
Therefore, to estimate event efficiency, it is necessary to study the scattering
angles of final-state particles and to compare them with the coverage angle.

\begin{figure}[t!]
  \centering
  \includegraphics[width=0.45\textwidth]{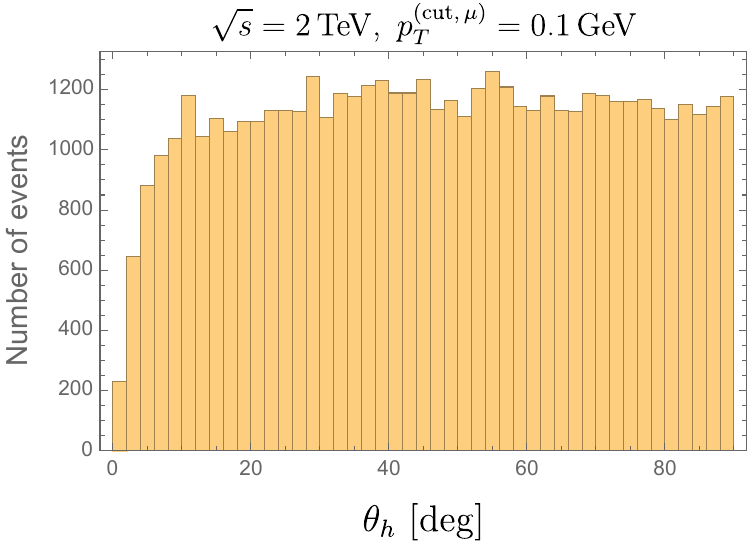} \hspace{3ex}
  \includegraphics[width=0.45\textwidth]{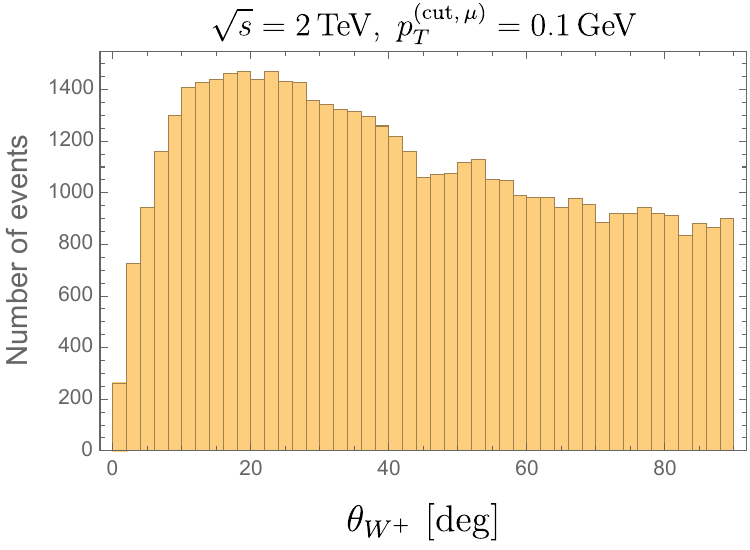} \\[3ex]
  \includegraphics[width=0.45\textwidth]{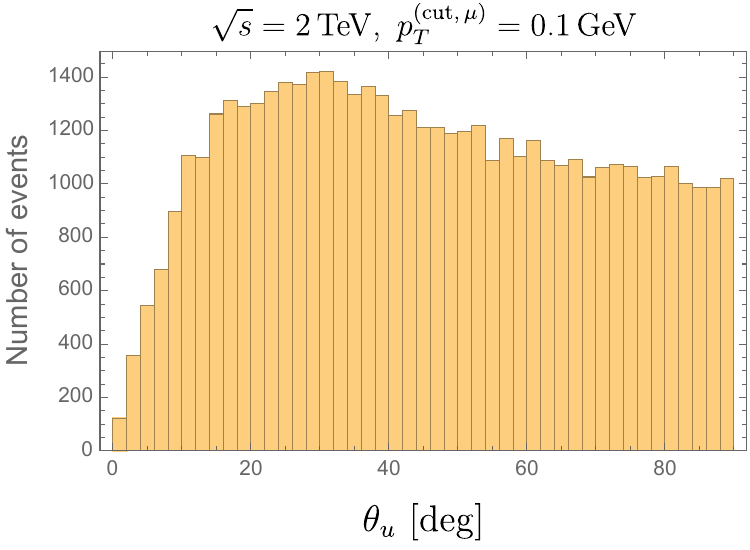} \hspace{3ex}
  \includegraphics[width=0.45\textwidth]{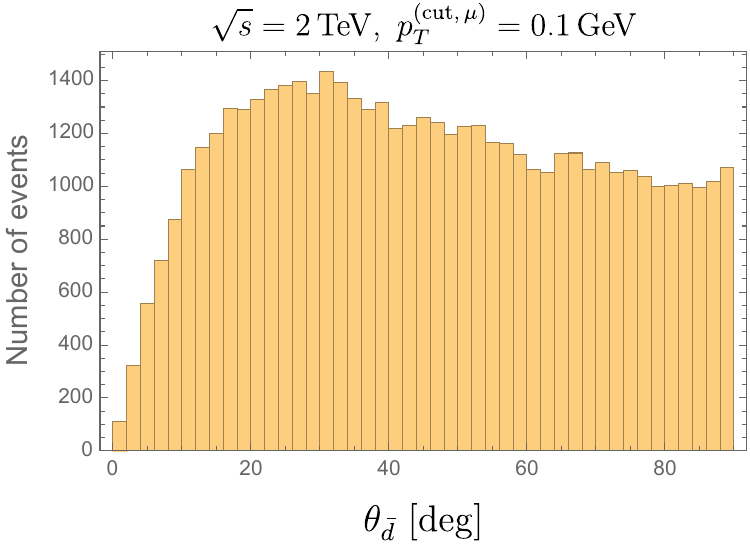} \\[3ex]
  \includegraphics[width=0.45\textwidth]{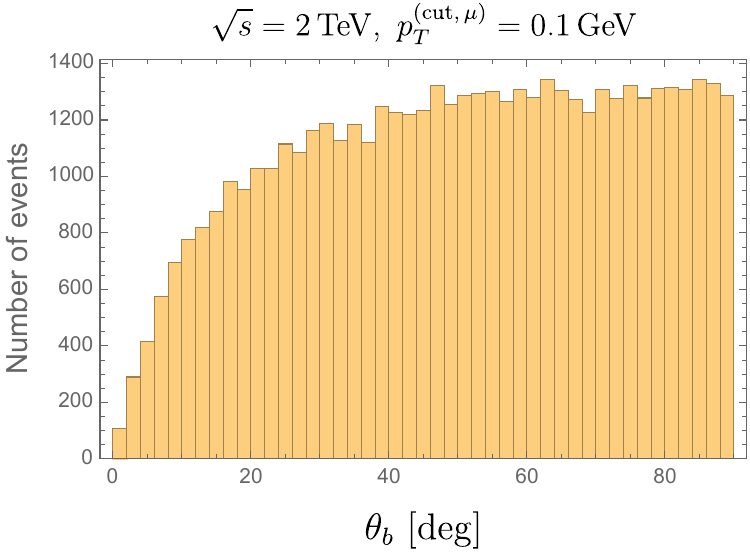} \hspace{3ex}
  \includegraphics[width=0.45\textwidth]{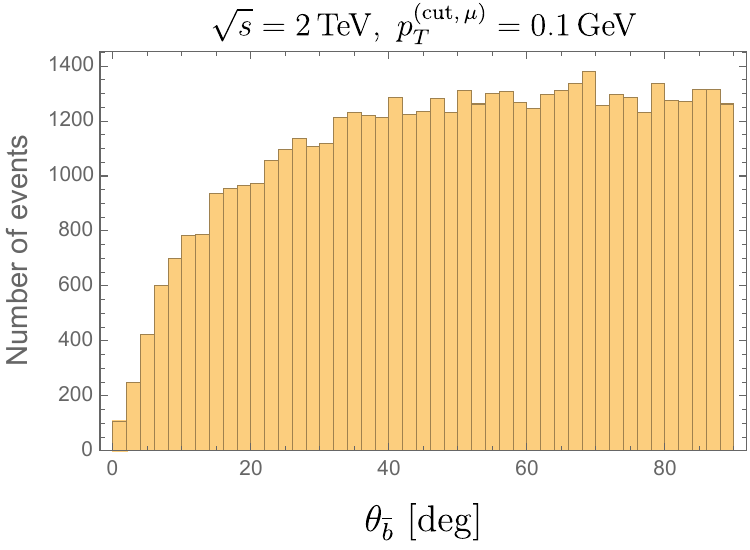}
  \caption{We show event histograms for the scattering angles of final-state particles in the $W$ boson fusion process with $\sqrt{s}=2$~TeV,
  where the calculation is performed by MadGraph with a minimum-$p_{\rm T}$ cut $0.1$~GeV.
  The number of the generated events is $5\times 10^4$.}
  \label{fig:MG_angle_hist_2TeV}
\end{figure}

\begin{figure}[t!]
  \centering
  \includegraphics[width=0.45\textwidth]{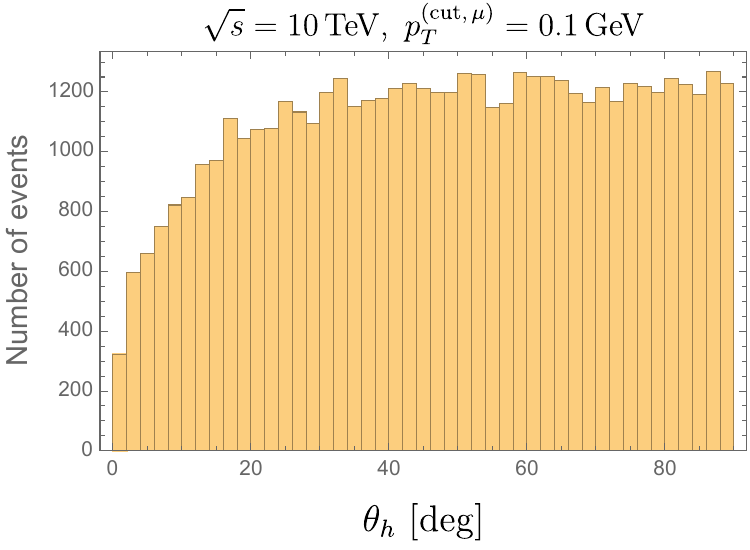} \hspace{3ex}
  \includegraphics[width=0.45\textwidth]{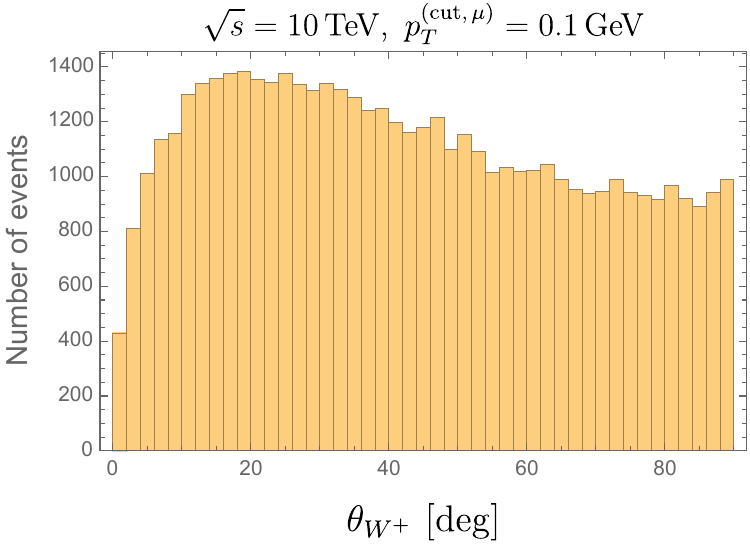} \\[3ex]
  \includegraphics[width=0.45\textwidth]{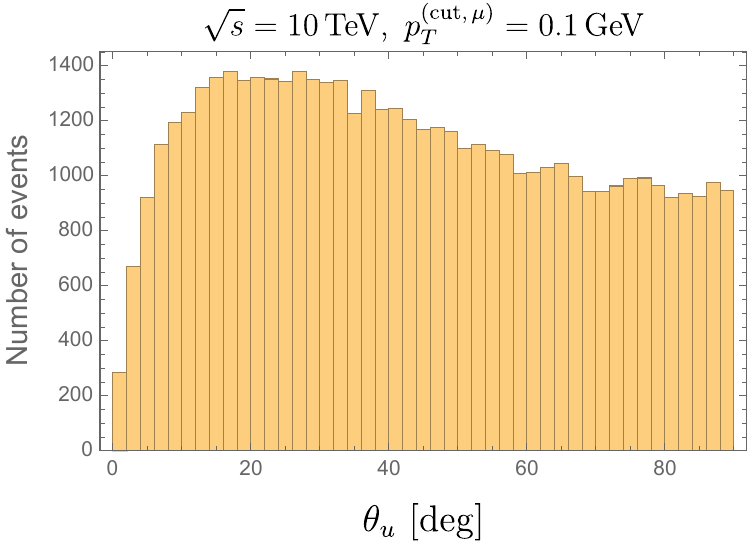} \hspace{3ex}
  \includegraphics[width=0.45\textwidth]{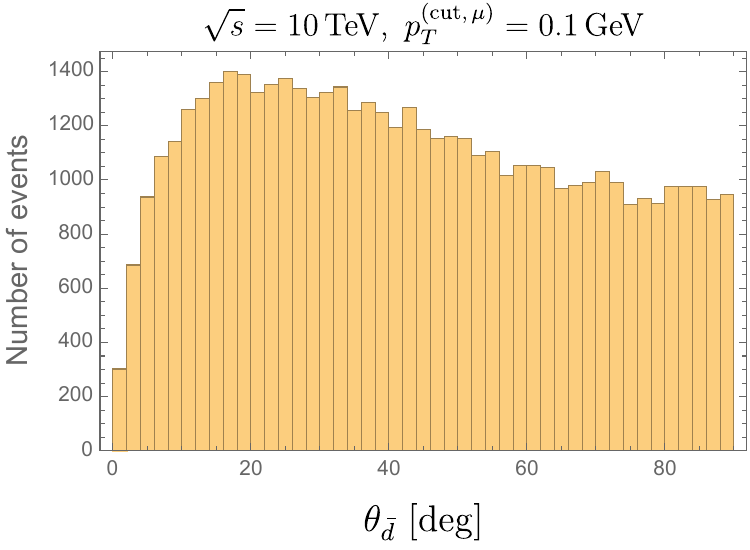} \\[3ex]
  \includegraphics[width=0.45\textwidth]{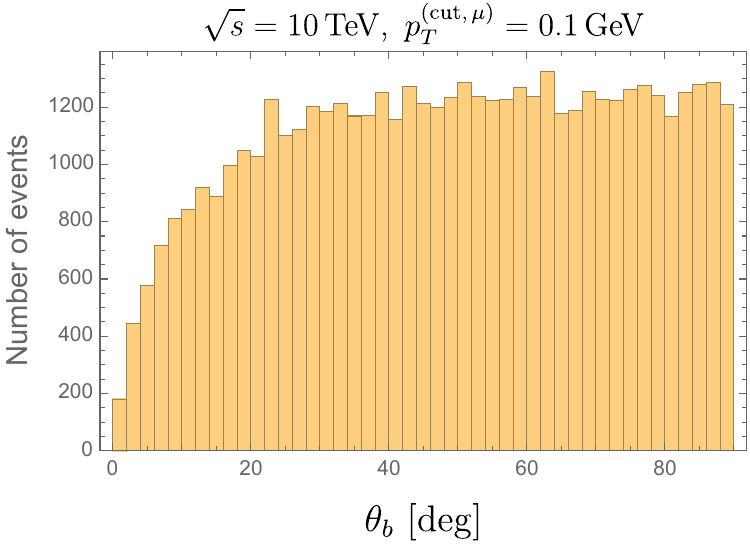} \hspace{3ex}
  \includegraphics[width=0.45\textwidth]{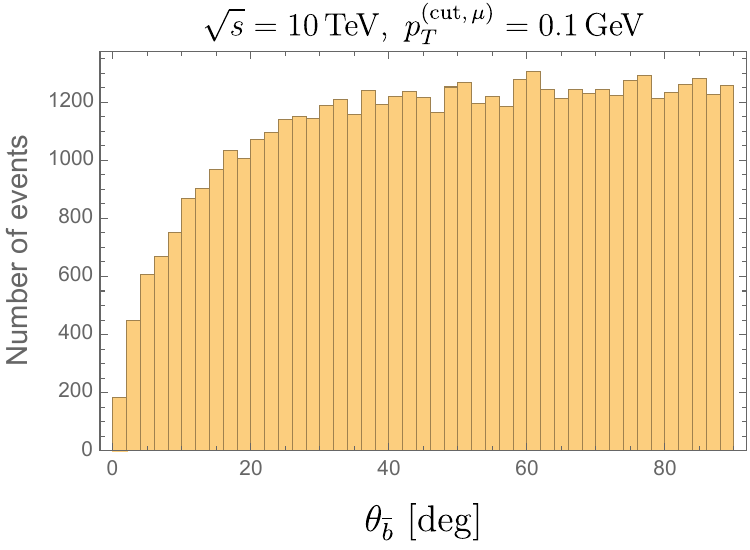}
  \caption{We show event histograms for scattering angles of final-state particles in the $W$ boson fusion process with $\sqrt{s}=10$~TeV,
  where the calculation is performed by MadGraph with a minimum-$p_{\rm T}$ cut $0.1$~GeV.
  The number of the generated events is $5\times 10^4$.}
  \label{fig:MG_angle_hist_10TeV}
\end{figure}

In the following, we present event histograms of the distribution of the scattering angles. As one of the main decay channels, we consider the produced $h$ and $W^+$ decaying into $b\bar b$, and light quarks ($u \bar d$), respectively.
We generate the events using MadGraph with $p_{\rm T} \geq 0.1$~GeV,
which corresponds to the high-$p_{\rm T}$ region of the previous sections.
Here, we neglect the low-$p_{\rm T}$ region 
because it does not have a significant number of events for this choice of lower-$p_{\rm T}$ cut.
We show in Figs.~\ref{fig:MG_angle_hist_2TeV} and \ref{fig:MG_angle_hist_10TeV}
the histograms for the intermediate $h$ and $W^+$ bosons and
their decay products $b$, $\bar{b}$, $u$, and $\bar{d}$ for $\sqrt{s}=2$~TeV and $10$~TeV, respectively. The histograms are at the parton level without smearing of the energies or momenta of the final-state particles.
Since a $\mu^+\mu^+$ collider is symmetric in the two initial-state particles,
we concentrate on the angles $0^\circ\leq \theta \leq 90^\circ$.

From these plots, we observe that most decay products have
sufficiently large scattering angles, or transverse momenta, and are hence visible. We therefore now estimate how many events can be caught by detectors.
Although there is not yet a concrete detector design for $\mu^+ \mu^+$
colliders, we assume the coverage of the 
hadron calorimeter to be around $10^\circ$~\cite{Mokhov:2011zzb,Mokhov:2011zzd}.
Thus, we find
that around $83\%$ of events can be caught by the detector, and reconstructing the $W$ and Higgs bosons is possible
for both $\sqrt{s}=2$~TeV and $10$~TeV. 
We expect this reconstruction to help to distinguish these signal events from background events. Further studies, such as detector simulation, will be performed elsewhere.


\section{Summary}
\label{sec:summary}

We studied Higgs boson production at $\mu^+ \mu^+$ colliders, where $W$ boson fusion is not possible at the leading order. This is because a $\mu^+$ cannot directly
emit the $W^-$ necessary to create the $W^+ W^-$ pair fusing to a Higgs boson. Nevertheless, we show that at high-energy $\mu^+ \mu^+$ colliders, a $\gamma$- and $Z$-mediated
$W$ boson fusion is possible, with its cross section growing as
$(\log s)^3$ with the center-of-mass energy $\sqrt{s}$. This means that at high energies, such colliders can produce almost as many
Higgs bosons as $\mu^+ \mu^-$ or $e^+ e^-$ colliders, since their production cross section only increases as $\log s$.

We calculated the cross section by carefully treating the divergence
associated with collinear emission of the intermediate photon from the antimuon.
Therein, we split the integration region of the transverse momentum $p_{\rm T}$ of the
antimuon into the sum of a low- and a high-$p_{\rm T}$ region. We then calculate the low-$p_{\rm T}$ part
using the equivalent photon approximation (EPA), and the high-$p_{\rm T}$ part directly using MadGraph.
EPA gives the leading-order value in a systematic expansion in
the maximal value of $p_{\rm T}$, and $Z$ boson contributions to the low-$p_{\rm T}$ region become important for $p_{\rm T} \gtrsim m_Z$. Therefore, we can control the systematic
uncertainties associated with this approximation by choosing a small
enough value $p_{\rm T}^{\rm (cut)}$ at which we split the low- and high-$p_{\rm T}$ regions. 
On the other hand, to avoid the collinear divergences or numerical instabilities from MadGraph, we must have $p_{\rm T}^{\rm (cut)} \gtrsim m_\mu$.
Thereby, we obtained a reliable result for the cross section by verifying
its independence on the $p_{\rm T}$ value at which we split the regions. This is well satisfied, e.g., for the choice $p_{\rm T}^{\rm (cut)} = \sqrt{m_\mu m_Z}$.

The cross section we obtained in this manner is indeed enhanced at high energy; note, however, that
the leading-logarithm formula that only includes $(\log s)^3$ terms seems to
overestimate it.
Nevertheless, even compared with the leading-order $W$ boson fusion process at
$\mu^+ \mu^-$ colliders, the cross section of the $\gamma$- and $Z$-mediated process with polarized beams can be as
large as half of the one for the leading-order process at 10~TeV energies. Therefore, while beam polarization capabilites are important, a $\mu^+
\mu^+$ collider is as good a Higgs boson factory as a $\mu^+ \mu^-$
collider.


\section*{Acknowledgements}

We would like to thank Koji Nakamura, Sayuka Kita and Mitsuhiro
Yoshida for useful discussions.
This work was in part supported by JSPS KAKENHI Grant Numbers
JP22K21350 (R.K., R.M. and S.O.), JP21H01086 (R.K.), JP19H00689 (R.K.), JP24KJ1157 (R.T.),
JP19K14711 (H.T.) and JP23K13110 (H.T.). H.T. is the Yukawa Research Fellow 
supported by Yukawa Memorial Foundation.
This work is also supported by the Deutsche Forschungsgemeinschaft under Germany's Excellence Strategy - EXC 2121 Quantum Universe - 390833306.


\appendix
\section{Equivalent photon approximation (EPA)}
\label{sec:iww_error}

Repeating a similar calculation to Ref.~\cite{Frixione:1993yw},
we derive the EPA formula applicable to the case 
where a $p_{\rm T}$ cut is applied. 
This section includes a review of Ref.~\cite{Frixione:1993yw}.   
Let us consider the photon-mediated process $\ell(p) + f(k) \to \ell(p') + X$, where
$\ell$, $f$, and $X$ denote a lepton, massless parton ($k^2=0$), and a generic final-state system, respectively.
The four-momenta of the leptons and the virtual photon are defined as
\begin{equation}
    p = E (1, 0, 0, \beta), \quad p' = E' (1, 0, \beta' \sin \theta, \beta' \cos \theta)
    \quad \text{and} \quad q = p - p'
\end{equation}
with
\begin{equation}
    \beta = \sqrt{1 - \frac{m_\ell^2}{E^2}} \quad \text{and} \quad \beta' = \sqrt{1 - \frac{m_\ell^2}{E'^2}} \, ,
\end{equation}
and $m_\ell$ the lepton mass.

The cross section is given by
\begin{equation}
    {\rm d} \sigma_{\rm EPA} = \frac{1}{8 k \cdot p}
    \frac{e^2 W^{\mu \nu} T_{\mu \nu}}{q^4} \frac{{\rm d}^3 p'}{(2\pi)^3 2 E'}
    = \frac{1}{8 k \cdot p} \frac{\alpha}{4 \pi} \frac{W^{\mu \nu} T_{\mu \nu}}{q^4}\,  {\rm d} q^2 {\rm d} x\, ,
    \label{eq:cross_section_WT}
\end{equation}
where $\alpha$ is the fine-structure constant, and the two independent kinematic variables, photon virtuality and longitudinal momentum fraction, are given by
\begin{equation}
    q^2 = 2 m_\ell^2 - 2 E E' (1 - \beta \beta' \cos \theta)
    \quad \text{and} \quad x = 1 - \frac{E' (1 + \beta' \cos \theta)}{E (1 + \beta)}\, ,
\end{equation}
respectively.
Furthermore, $x$ satisfies
$ x=1-\frac{k\cdot p'}{k\cdot p}=\frac{k\cdot q}{k\cdot p}$.
The leptonic tensor $T_{\mu \nu}$ and 
the hadronic tensor $W^{\mu \nu}$ of the electromagnetic current are given by
\begin{equation}
    T_{\mu \nu} = 4 \left( \frac{1}{2} q^2 g_{\mu\nu} + p_\mu p'_\nu + p_\nu p'_\mu \right)
\end{equation}
and
\begin{equation}
    W^{\mu\nu} = W_1 (q^2, k \cdot q) \left(- g^{\mu\nu} + \frac{q^\mu q^\nu}{q^2} \right)
    - \frac{q^2 W_2(q^2, k\cdot q)}{(k \cdot q)^2} \left(k^\mu - \frac{k \cdot q}{q^2} q^\mu \right) \!\! \left(k^\nu-\frac{k \cdot q}{q^2} q^\nu \right).
\end{equation}
Contracting them then yields
\begin{equation}
    \begin{split}
        W^{\mu\nu} T_{\mu\nu}
        &=-4 \left[ 2 m_\ell^2 W_1(q^2, k \cdot q) + q^2 \left(W_1(q^2, k \cdot q) + \frac{2 (1-x)}{x^2} W_2(q^2, k \cdot q) \right) \right] \\
        &\simeq -4 W_1(0, k\cdot q) \left[2 m_\ell^2 + q^2 \frac{1 + (1-x)^2}{x^2} \right],
    \end{split}
    \label{eq:contraction_WT}
\end{equation}
where we use
\begin{equation}
    W_1(q^2, k\cdot q) = W_1(0, k\cdot q) + \mathcal{O}(q^2) \quad \text{and} \quad
    W_2(q^2, k\cdot q) = W_1(0, k\cdot q) + \mathcal{O}(q^2)\, ,
\end{equation}
obtained by requiring that $W^{\mu \nu}$ be analytic in $q^2$ as $q^2 \to 0$.

Therefore, the cross section can be approximated by 
\begin{equation}
    {\rm d} \sigma_{\rm EPA} = f_{\gamma/\ell}(x) \sigma_{\gamma f}(q, k) {\rm d}x\, ,
\end{equation}
with~\cite{Frixione:1993yw}
\begin{equation}
    f_{\gamma/\ell}(x) = \frac{\alpha}{2 \pi} 
    \left[ 2 m_\ell^2 x \left( \frac{1}{q_{\rm max}^2}-\frac{1}{q_{\rm min}^2} \right) + \frac{1 + (1-x)^2}{x} \log \frac{q_{\rm min}^2}{q_{\rm max}^2} \right] \label{eq:IWWPDF}
\end{equation}
and 
\begin{equation}
    \sigma_{\gamma f}(q,k)=\frac{W_1(0, k\cdot q)}{4 k \cdot q}\, .
\end{equation}
We obtain these expressions by integrating Eq.~\eqref{eq:cross_section_WT} with Eq.~\eqref{eq:contraction_WT} over $q^2$ from $q_{\rm min}^2$ to $q_{\rm max}^2$.

Ref.~\cite{Frixione:1993yw} further presents the formula for the case
where a $\theta$ cut is applied, with $0 \leq \theta < \theta_c$ with $\theta_c \ll 1$.
In this case, the range of $q^2$ reads~\cite{Frixione:1993yw}
\begin{equation}
    q_{\rm max}^2 = -\frac{m_\ell^2 x^2}{1-x} \quad \text{and} \quad
    q_{\rm min}^2 = -\frac{m_\ell^2 x^2}{1-x} - E^2 (1-x) \theta_c^2\, .
    \label{eq:IWWPDF_qmaxmin_theta}
\end{equation}
Substituting these results into Eq.~\eqref{eq:IWWPDF}, one obtains $f_{\gamma/\ell}(x)$
in the $\theta$-cut scheme.

Instead of a $\theta$-cut, we give $f_{\gamma/\ell}(x)$ for the case where a $p_{\rm T}$ cut is applied, with
$0 \leq p_{\rm T} < p_{\rm T}^{\rm (cut)}$.
In particular, we find a simple relation between $p_{\rm T}^2$ and $q^2$:
\begin{equation}
    p_{\rm T}^2 = - q^2 (1-x) - m_\ell^2 x^2\, ,
\end{equation}
as seen from $p_{\rm T} = E' \beta' \sin \theta$. This is especially useful, since it does not depend on $E$, in contrast to the case with $\theta$.
Therefore, the range of $q^2$ reads
\begin{equation}
    q_{\rm max}^2 = -\frac{m_\ell^2 x^2}{1-x} \quad \text{and} \quad
    q_{\rm min}^2 = -\frac{1}{1-x} \left( \left( p_{\rm T}^{\rm (cut)} \right)^2 + m_\ell^2 x^2 \right).
    \label{eq:IWWPDF_qmaxmin_pT}
\end{equation}
Substituting these results into Eq.~\eqref{eq:IWWPDF}, we obtain $f_{\gamma/\ell}(x)$
in the $p_{\rm T}$ cut scheme, as shown in Eq.~\eqref{eq:iww}.

Next, we discuss the uncertainty of the EPA formula. The EPA
calculation utilizes the leading term of the expansion of $W_i(q^2, k
\cdot k)$ in $q^2$:
\begin{equation}
    W_i(q^2, k \cdot q) = \sum_{n \geq 0} A_i^{(n)}(k \cdot q) \left( \frac{q^2}{(k \cdot q)} \right)^n
\end{equation}
for $i = 1$, $2$.
Therefore, we study the impact of higher-order terms by again following 
the corresponding calculation in Ref.~\cite{Frixione:1993yw}.
Due to 
\begin{equation}
    \begin{split}
        W^{\mu\nu} T_{\mu\nu} = &- 4 W_1(0, k \cdot q) \left[ 2 m_\ell^2 + q^2 \frac{1+(1-x)^2}{x^2} \right]
        - 8 m_\ell^2 \left(A^{(1)}_1 \frac{q^2}{k\cdot q} + A^{(2)}_1 \frac{q^4}{(k \cdot q)^2}\right) \\[3pt]
        &- 4 q^2 \left( A^{(1)}_1 \frac{q^2}{k\cdot q} + \frac{2 (1-x)}{x^2} A^{(1)}_2\frac{q^2}{k \cdot q} \right) + \mathcal{O}(q^6)\, ,
    \end{split}
\end{equation}
the error caused by the higher-order terms is given by
\begin{equation}
    \label{xserror}
    \begin{split}
        &\Delta {\rm d} \sigma_{\rm EPA} \\
        &\simeq - \frac{\alpha}{2\pi} \frac{1}{4 k \cdot p} \left\{ \frac{2 m_\ell^2 A_1^{(1)}}{k \cdot q} \frac{1}{q^2}
        + \frac{2 m_\ell^2 A_1^{(2)}}{(k \cdot q)^2} + \frac{A_1^{(1)}}{k \cdot q} + \frac{2 (1-x)}{x^2} \frac{A_2^{(1)}}{k \cdot q} \right\} {\rm d}q^2 {\rm d}x \\[5pt]
        &\simeq - \frac{\alpha}{2\pi} \left\{ \frac{m_\ell^2}{s} \frac{A_1^{(1)}}{k \cdot q} \log{\frac{q^2_{\rm max}}{q^2_{\rm min}}}
        + \left[\frac{2 m_\ell^2}{s} \frac{A_1^{(2)}}{k \cdot q} \frac{1}{x} + \frac{1}{2} \frac{A_1^{(1)}}{k \cdot q} + \frac{A_2^{(1)}}{k \cdot q} \frac{1-x}{x^2} \right] \frac{q^2_{\rm max}-q^2_{\rm min}}{s} \right\} {\rm d}x\, .
    \end{split}
\end{equation}
In the second equality, we performed the $q^2$-integral, then used the relation
\begin{equation}
    k \cdot q = x k \cdot p = \frac{x s}{2} + \mathcal{O} \left( \frac{m_\ell^2}{s} \right),
\end{equation}
and finally combined $1/(k \cdot q)$ with $A_i^{(n)}$ to take 
\begin{equation}
    \frac{A_i^{(n)}}{k \cdot q} \sim \mathcal{O}(1) \times \sigma_{\gamma f}.
\end{equation}
This identification is analogous to what we did for the $n=0$ case.
Noting that
\begin{equation}
    q^2_{\rm max} - q^2_{\rm min} =
    \begin{cases}
        &E^2(1-x) \theta_c^2 \quad  \text{for a $\theta$ cut}, \\
        &\frac{1}{1-x} \big(p_{\rm T}^{\rm (cut)}\big)^2  \quad \,  \text{for a $p_{\rm T}$ cut,}
    \end{cases}
\end{equation}
and also 
\begin{equation}
    \int_{x_{\rm min}} \frac{{\rm d}x}{x^2} \sim \frac{1}{x_{\rm min}}\, ,
\end{equation}
where $x_{\rm min}$ is given by $s_{\rm min}/s$,
we see that the dominant error
stems from the last term inside the square brackets of Eq.~\eqref{xserror}, and we thus find:
\begin{equation}
    \frac{\Delta \sigma_{\rm EPA} \big|_{p_{\rm T} < p_{\rm T}^{\rm (cut)}}}
    {\sigma_{\rm EPA} \big|_{p_{\rm T} < p_{\rm T}^{\rm (cut)}}} \sim
    \begin{cases} 
        \mathcal{O}(E^2 \theta_c^2/s_{\rm min}) \quad \quad \, \,  \text{for a $\theta$ cut,} \\
        \mathcal{O}\big(\big(p_{\rm T}^{\rm (cut)}\big)^2/s_{\rm min}\big)  \quad  \text{for a $p_{\rm T}$ cut}\, .
    \end{cases}
    \label{eq:iww_relative_error}
\end{equation}


\section{Error in the massless lepton approximation}
\label{sec:mass_error}

We now discuss the error in the calculation with MadGraph for the high-$p_{\rm T}$ cross section
in Eq.~\eqref{eq:CS-Overview_Split_Cross_IWW_MG} when the muon mass is neglected.
For this, we split $\sigma_{\rm MG}\big|_{p_{\rm T} \geq p_{\rm T}^{\rm (cut)}}$ as
\begin{equation}
    \sigma_{\rm MG}\Big|_{p_{\rm T} \geq p_{\rm T}^{\rm (cut)}}
    =\sigma_{\rm MG}\Big|_{p_{\rm T}^* \geq p_{\rm T} \geq p_{\rm T}^{\rm (cut)}}
    +\sigma_{\rm MG}\Big|_{p_{\rm T} > p_{\rm T}^*}
\end{equation}
by using an additional cut $p_{\rm T}^*$. 
The dominant error caused by the massless approximation should stem 
from the first cross section of the sum, where $p_{\rm T}$ is smaller.
We then set the new cut $p_{\rm T}^*$ such that the EPA formula is also valid for
the cross section with $p_{\rm T}^* \geq p_{\rm T} \geq p_{\rm T}^{\rm (cut)}$. 
(The plateaus we observe in Fig.~\ref{fig:HPR-Summed_Data_vs_pT}
indicate that such a $p_{\rm T}^*$ exists.)
Thus, we can estimate the error by
\begin{eqnarray}
    \label{eq:sigmaMGerror}
    \Delta \sigma_{\rm MG} \Big|_{p_{\rm T} \geq p_{\rm T}^{\rm (cut)}}
    &\simeq& \sigma_{\rm MG}\Big|_{p_{\rm T}^* \geq p_{\rm T} \geq p_{\rm T}^{\rm (cut)}}
    - \sigma_{\rm MG}\Big|_{p_{\rm T}^* \geq p_{\rm T} \geq p_{\rm T}^{\rm (cut)}, \, m_\mu^2 = 0} \nonumber \\ [3pt]
    &\simeq& \sigma_{\rm EPA}\Big|_{p_{\rm T}^* \geq p_{\rm T} \geq p_{\rm T}^{\rm (cut)}}
    - \sigma_{\rm EPA}\Big|_{p_{\rm T}^* \geq p_{\rm T} \geq p_{\rm T}^{\rm (cut)}, \, m_\mu^2 = 0} \nonumber \\[3pt]
    &=& \int {\rm d}x \left[ f_{\gamma/\mu}^{p_{\rm T}^{\rm (cut)}, \, p_{\rm T}^*} (x) - f_{\gamma/\mu}^{p_{\rm T}^{\rm (cut)}, \, p_{\rm T}^*} (x) \Big|_{m_\mu^2=0} \right] \sigma_{\gamma\mu}(xs)\, ,
\end{eqnarray}
where the function $f_{\gamma/\mu}^{p_{\rm T}^{\rm (cut)}, \, p_{\rm T}^*} (x)$ denotes
the EPA parton distribution function in Eq.~\eqref{eq:IWWPDF} with
\begin{equation}
    q_{\rm min}^2 = - \frac{1}{1-x} \left( p_{\rm T}^{* \, 2} + m_\mu^2 x^2 \right)
    \quad {\rm and} \quad
    q_{\rm max}^2 = - \frac{1}{1-x} \left( \big(p_{\rm T}^{\rm (cut)}\big)^2 + m_\mu^2 x^2 \right),
\end{equation}
and the function $f_{\gamma/\mu}^{p_{\rm T}^{\rm (cut)}, \, p_{\rm T}^*} (x) \Big|_{m_\mu^2=0}$ denotes the analogous expression with $m_\mu \to 0$.

Assuming $m_\mu^2 \ll \big(p_{\rm T}^{\rm (cut)}\big)^2 \, \ll p_{\rm T}^{* \, 2}$, the difference of these functions is
\begin{eqnarray}
    &&f_{\gamma/\mu}^{p_{\rm T}^{\rm (cut)}, \, p_{\rm T}^*} (x) - f_{\gamma/\mu}^{p_{\rm T}^{\rm (cut)}, \, p_{\rm T}^*} (x) \Big|_{m_\mu^2=0} \nonumber \\
    &=& \frac{\alpha}{2\pi} 
    \left[ 2 m_\mu^2 x \left( \frac{1-x}{m_\mu^2 x^2+{p_{\rm T}^*}^2} -\frac{1-x}{m_\mu^2 x^2+\big(p_{\rm T}^{\rm  (cut)}\big)^2}\right) 
    +\frac{1+(1-x)^2}{x} \log{ \frac{1+m_\mu^2 x^2/{p_{\rm T}^*}^2}{1+m_\mu^2 x^2/\big(p_{\rm T}^{\rm (cut)}\big)^2}} \right] \nonumber \\
    &\simeq& - \frac{m_\mu^2}{\big( p_{\rm T}^{\rm (cut)} \big)^2} \frac{\alpha}{2 \pi} x (2-x)^2\, ,
    \label{eq:diffPDF}
\end{eqnarray}
and therefore the error size is controlled by $m_\mu^2/\big(p_{\rm T}^{\rm (cut)}\big)^2$.
Furthermore, we observe that Eq.~\eqref{eq:diffPDF} is not singular at $x=0$,
and loses the logarithmic factor
present in $f^{p_{\rm T}^{\rm (cut)}}_{\gamma/\mu} \!\!\! (x)$.
This indicates that Eq.~\eqref{eq:sigmaMGerror} 
has two fewer logarithmic factors compared with $\sigma \big|_{p_{\rm T} \geq p_{\rm T}^{\rm (cut)}}$.
Noting also that the infrared cutoff for $\sigma \big|_{p_{\rm T} \geq p_{\rm T}^{\rm (cut)}}$
is $p_{\rm T}^{\rm (cut)}$, rather than $m_\mu$,
we hence estimate the relative uncertainty by
\begin{equation}
    \frac{\Delta \sigma_{\rm MG}\Big|_{p_{\rm T} \geq p_{\rm T}^{\rm (cut)}}}{\sigma \big|_{p_{\rm T} \geq p_{\rm T}^{\rm (cut)}}} 
    \sim \frac{m_\mu^2}{\left( p_{\rm T}^{\rm (cut)} \right)^2}
    \left[ \log \frac{s}{\left( p_{\rm T}^{\rm (cut)} \right)^2}
    \log{\frac{s}{s_{\rm min}}} - \frac{2}{3} \left(\log{\frac{s}{s_{\rm min}}}\right)^2 \right]^{-1}.
    \label{eq:mg_massless_error}
\end{equation}

\section{Semi-automatic MadGraph implementation of \texorpdfstring{$p_{\rm T}^{\rm (cut)}$}{pTcut} calculation}
\label{sec:madgraph}

As previously mentioned, we also implemented a semi-automatic calculation of the $p_{\rm T}^{\rm (cut)}$ scheme for convenience. Here, semi-automatic means that both the low- and high-$p_{\rm T}$ cross sections can be calculated in MadGraph directly; however, the two parts still need to be summed by the user. In particular, we implement the calculation of the low-$p_{\rm T}$ cross section using EPA, since the high-$p_{\rm T}$ part can already be computed directly. In this appendix, we therefore discuss how the calculation is implemented, how to use it, and its accuracy.

First, we briefly mention how MadGraph calculates EPA, or IWW, via the \verb|iww| setting in the run card. With this setting, MadGraph uses the parton distribution function formula of Eq.~\eqref{eq:IWWPDF} with $q_{\rm max}^2 = - m_\ell^2 x^2 / (1 - x)$, as given in Eq.~\eqref{eq:IWWPDF_qmaxmin_theta}; we also note here that the lepton mass in $q_{\rm max}^2$ is hard-coded, and specified through the collider type set in the run card of the respective run.
However, $q_{\rm min}^2$ is not given by a fixed expression, but is instead provided to the parton distribution function as external input, together with the momentum fraction $x$. 
In particular, $q_{\rm min}^2$ takes on the role of (squared) factorization scale, which is set either by a fixed scale for all events or a dynamical scale determined on an event-by-event basis, depending on the user specification in the run card. This is because the factorization scale is the largest scale included in the parton distribution function description, whose square is precisely the maximum absolute value of the photon virtuality, $\left| q_{\rm min}^2 \right|$.
Thus, based on the dynamical scale choice specified by the user, $q_{\rm min}^2$ is calculated by MadGraph for each event, and then passed on to the IWW parton distribution function. For instance, the setting \verb|4| sets the dynamical (factorization) scale to the partonic center-of-mass energy, which is a common choice, and thus $\left| q_{\rm min}^2 \right| = x s$. Note that this is calculated using the four-momenta of the incoming partons of the respective event; and if applicable, $x$ may be a product of momentum fractions of the incoming partons.

Since MadGraph supports user-defined dynamical scales, we implement the parton distribution function of Eq.~\eqref{eq:iww} through such a custom dynamical scale, which can then be imported to MadGraph and used for the low-$p_{\rm T}$, EPA calculation.
The implementation is partly based on the original MadGraph code, and works in the following way: First, we import the collider information, in particular the beam energies and beam types. For this implementation, the beams are restricted to electrons, muons, and their respective anti-particles. Second, we import the kinematical information of the respective parton event and define the quantities necessary for the calculation. Therein, the value of $p_{\rm T}^{\rm (cut)}$ is hard-coded, and thus needs to be changed by hand depending on which value is required for the process of interest.
Third, we calculate and set the pertinent variables from the kinematical data. In particular, we set the lepton mass based on the information from the run card, and then calculate the momentum fraction $x$ using
\begin{equation}
    \label{eq:mg_iww_x}
    x = \frac{\hat{s}}{s} + \mathcal{O} \left( \frac{m_{\rm beam\, 1, 2}^2}{s} \right)
    = \frac{\big(p_{\rm parton\, 1} + p_{\rm parton\, 2}\big)^2}{\big(E_{\rm beam\, 1} + E_{\rm beam\, 2}\big)^2}
    + \mathcal{O} \left( \frac{m_{\rm beam\, 1, 2}^2}{s} \right),
\end{equation}
where $\sqrt{s}$ is the center-of-mass energy of the collider, $\sqrt{\hat{s}}$ that of the partonic process, $m_{\rm beam\, 1, 2}$ are the masses of the beam particles set in the run card, $p_{\rm parton\, 1, 2}$ are the four-momenta of the incoming partons, and $E_{\rm beam\, 1, 2}$ are the respective beam energies. Here, we used that $\hat{s}$ is determined by the incoming partons' four-momenta and is equal to the product $x s$, up to corrections of $\mathcal{O} (m_{\rm beam\, 1, 2}^2 /s)$.
Lastly, we calculate the dynamical scale via Eq.~\eqref{eq:IWWPDF_qmaxmin_pT} as
\begin{equation}
    {\rm (dynamical\, scale)} = \sqrt{\left| q_{\rm min}^2 \right|} = \sqrt{\frac{1}{1-x}\left(\left( p_{\rm T}^{\rm (cut)} \right)^2 + m_\ell^2 x^2\right)}\, ,
\end{equation}
and return the result, which will then be used in the \verb|iww| parton distribution function.

To use this implementation in practice, the following steps are necessary. First, save the code as \verb|.f| (Fortran) text file on your machine, and copy the file path. Second, adjust the $p_{\rm T}^{\rm (cut)}$ value in GeV according your process of interest and save the changes; the default setting is the geometric mean of the muon and $Z$ boson mass, $\sqrt{m_\mu m_Z} \simeq 3.10$~GeV, since this was the pertinent value for our considerations. The corresponding variable in the file is denoted as \verb|pt_cut_iww| and marked by a box around it for ease of use. Third, run MadGraph as usual, set the photon-emitting beam as $\pm$\verb|3| for $e^\mp$ or $\pm$\verb|4| for $\mu^\mp$, and the parton distribution function for the photon to \verb|iww|. Fourth, set the \verb|dynamical_scale_choice| parameter to \verb|0|, which means that a custom dynamical scale is to be used. Note, in particular, that no fixed factorization scale should be set. Lastly, set the parameter \verb|custom_fcts| to the file path of the \verb|.f| file containing the code to calculate $\sqrt{\left| q_{\rm min}^2 \right|}$. After this, the run can be continued in the usual manner.

Note that to obtain the correct low-$p_{\rm T}$ cross section for the case of identical beam particles, the obtained result needs to be multiplied by a factor of two, as is also the case for a manual convolution.

In terms of accuracy, the results thereby obtained by the MadGraph implementation of the low-$p_{\rm T}$ cross section are consistently about $5\%$ smaller on average, when compared with the manual calculation for our Higgs production process at center-of-mass energies up to $30$~TeV. The deviation is a bit larger for the lower center-of-mass energies, and a bit smaller for the higher ones, with a respective $\mathcal{O}(1)\%$ change in both directions. Nevertheless, we observe an approximate plateau for similar $p_{\rm T}^{\rm (cut)}$ values as in the manual calculation, irrespective of beam polarizations and energies.
Since the size of the deviation of the cross section is largely independent of beam polarizations and center-of-mass energies, we assume that it is a systematic error. Numerically, it can be mitigated by adjusting the \verb|scalefact| parameter in the run card to the value $\sim 1.3$, which modifies the dynamical scale as $\,{\rm (new\, dynamical\, scale)} = \verb|scalefact| \times {\rm (old\, dynamical\, scale)}\,$. The results obtained from this modification are within about $\pm 0.4\%$ of the manual calculation, across center-of-mass energies and polarizations.

Finally, note that due to the calculation of $x$ as $\hat{s} / s$, and since the dynamical scale is calculated based on Eq.~\eqref{eq:IWWPDF_qmaxmin_pT} specifically for the photon, we recommend using partonic processes where a parton distribution function is only used for the photon.


\bibliographystyle{JHEP} 
\bibliography{sn-bibliography}

\end{document}